\documentclass{aa}

\usepackage{amsmath}
\usepackage{amsfonts}
\usepackage{amssymb}

\usepackage{mathenv}

\usepackage{graphicx}

\usepackage{txfonts}

\usepackage{natbib}
\bibpunct{(}{)}{;}{a}{}{,}

\defcitealias{bra2008}{Paper I}

\begin{document}
%
%
%
%
\title{Relativistic tidal compressions of a star by a massive black hole}
\author{M. Brassart \and J.-P. Luminet}
\offprints{M. Brassart}
\institute{Laboratoire Univers et Th\'eories,
           Observatoire de Paris,
	   CNRS,
	   Universit\'e Paris Diderot,
	   5 place Jules Janssen, \\
	   F-92190 Meudon, France \\
           \email{[matthieu.brassart;jean-pierre.luminet]@obspm.fr}
           }
\date{Received date; accepted date}
%
\abstract
%
%
{}
%
%
{
We investigate the stellar pancake mechanism during which a solar-type star is tidally flattened within its orbital plane passing close to a $10^{6} \, M_{\odot}$ black hole.
We simulate the relativistic orthogonal compression process and follow the associated shock waves formation.
}
%
%
{
We consider a one-dimensional hydrodynamical stellar model moving in the relativistic gravitational field of a non-rotating black hole.
The model is numerically solved using a Godunov-type shock-capturing source-splitting method in order to correctly reproduce the shock waves profiles.
}
%
%
{
Simulations confirm that the space-time curvature can induce several successive orthogonal compressions of the star which give rise to several strong shock waves.
The shock waves finally escape from the star and repeatedly heat up the stellar surface to high energy values.
Such a shock-heating could interestingly provide a direct observational signature of strongly disruptive star - black hole encounters through the emission of hard $X$ or soft $\gamma$ -ray bursts.
Timescales and energies of such a process are consistent with some observed events such as GRB 970815.
}
%
%
{}
%
\maketitle
%
%
%
%
\section{Introduction}
It has long been underlined that stars orbiting closely enough to massive black holes could be tidally flattened into a transient \textit{pancake}-shape configuration until full disruption \citep{car1982,car1983}.
Strongest compressions have been predicted to trigger  thermonuclear explosion in the core of small main-sequence stars grazing black holes from thousands up to millions solar masses  \citep{pic1985,lum1989a}.
The presence of specific proton-enriched chemical elements near galactic centres may be compatible with the stellar pancakes nucleosynthesis \citep{lum1990}.

More recently, high-resolution hydrodynamical simulations have shown that the tidal compression could independently give rise to strong shock waves within the stellar matter \citep{kob2004,bra2008,gui2009}. 
Propagating outwards, shock waves may then be able to quickly heat up the stellar surface and lead to the emission of a new type of hard $X$ or soft $\gamma$ -ray bursts.

Aftermath of tidal disruptions of stars by massive black holes have already been detected from nearby otherwise quiescent galactic cores through giant-amplitude persistent $X$-$UV$ flares \citep[e.g.][]{bad1996,gru1999,kom1999,gre2000,kom2002,hal2004,kom2004,gez2006,esq2007,gez2008,cap2009}.
These flares have arised from the long term evolution when part of the liberated stellar gas was being accreted by the central black holes.
However, the very short-timescale high-energy flares due to the initial compression of the star could interestingly allow to catch disruptions from their beginning.

In our previous article \citep[][hereafter \citetalias{bra2008}]{bra2008}, we have simulated the tidal compression orthogonal to the orbital plane using a one-dimensional Godunov-type shock-capturing hydrodynamical model, and assuming that the star evolved within a Newtonian external gravitational field.

In this article, we naturally extend the model to the general relativistic case of a non-rotating Schwarzschild black hole and present some complementary results.

In the following, relativistic equations are expressed in geometrized units $c=G=1$, greek and latin indices run respectively from 0 to 3 and 1 to 3.
%
%
%
%
\section{Basic equations}
\subsection{Orbital equations of motion and relativistic tidal field}
We consider the Schwarzschild coordinates in which the black hole of mass $M_{\bullet}$ is described by the usual spherically symmetric line element
\begin{eqnarray}
ds^{2} = - \left( 1 - \frac{2 M_{\bullet}}{r} \right) dt^{2} 
         &+& \left( 1 - \frac{2 M_{\bullet}}{r} \right)^{-1} dr^{2} \nonumber \\
         &+& \, r^{2} (d\theta^{2} + sin^{2}\theta \, d\phi^{2}). \label{eq_e_01}
\end{eqnarray}
The motion of the star's centre of mass follows a timelike geodesic of proper time $\tau$ defined by \citep[e.g.][]{cha1983}
\begin{eqnarray}
\dot{t}      & = & E \left( 1 - \frac{2 M_{\bullet}}{r} \right)^{-1}, \label{eq_e_02} \\
\dot{r}^{2}  & = & E^{2} - 1 + \frac{2 M_{\bullet}}{r} - \frac{L^{2}}{r^{2}} + \frac{2 M_{\bullet} L^{2}}{r^{3}}, \label{eq_e_03} \\
\dot{\theta} & = & 0, \label{eq_e_04} \\
\dot{\phi}   & = & \frac{L}{r^{2}}, \label{eq_e_05}
\end{eqnarray}
with $\left(\, \dot{} \, \right)$ the derivative with respect to $\tau$.
Such an orbit lies in the equatorial plane $\theta=\pi / 2$ and entirely depends on both constants of motion $E$ and $L$, respectively the specific total energy and the specific orbital angular momentum of the star.

The tidal field applied within the star is characterized by the relative acceleration between infinitesimal matter elements viewed as a collection of test particles.
Let $u^{\mu}$ be the four-velocity of the star's centre of mass, and $k^{\mu}$ the four-separation with a nearby test particle moving along an adjacent geodesic such that $u^{\mu}k_{\mu}=0$.
Their relative acceleration obeys the geodesic deviation equation \citep[e.g.][]{mis1973}
\begin{eqnarray}
(\mathbf{\nabla}_{\boldsymbol{u}} \mathbf{\nabla}_{\boldsymbol{u}} \boldsymbol{k})^{\sigma} & = & u^{\mu} \, u^{\nu} \, \nabla_{\mu} \nabla_{\nu} \, k^{\sigma} \label{eq_e_09} \\
& = & - R^{\sigma}_{\mu \nu \rho} \, u^{\mu} \, k^{\nu} \, u^{\rho}, \label{eq_e_10}
\end{eqnarray}
with $\nabla$ the covariant derivative and $R^{\sigma}_{\mu \nu \rho}$ the Riemann curvature tensor.
Let $w^{\mu}_{(\alpha)}$ be an orthonormal tetrad locally defined at the star's centre of mass, and parallely propagated along the geodesic motion such that
\begin{eqnarray}
w^{\mu}_{(0)} & = & u^{\mu}, \label{eq_e_11} \\
w^{\mu}_{(\alpha)} \, g_{\mu \nu} \, w^{\nu}_{(\beta)} & = & \eta_{\alpha \beta}, \label{eq_e_12}
\end{eqnarray}
where the indice in brackets refers to a vector of the tetrad, $g_{\mu \nu}$ and $\eta_{\mu \nu}=\mathrm{diag}(-1,1,1,1)$ are the metric tensors respectively of the space-time around the black hole and of the flat space-time.
For a comoving observer at the star's centre of mass, labelling the infinitesimal nearby matter elements by the three-position
\begin{equation}  
z^{i} = w^{(i)}_{\mu} k^{\mu}, \label{eq_e_13}
\end{equation}
the tidal acceleration (\ref{eq_e_09}) can thus be written in the Newtonian-type form \citep{pir1956}
\begin{equation}
\ddot{z}^{i} = z^{j} \, C^{i}_{j}, \label{eq_e_14}
\end{equation}
where the tidal tensor is defined in terms of the Riemann curvature tensor as
\begin{equation}
C_{ij} = - R_{\mu \nu \rho \sigma} \, w^{\mu}_{(0)} \, w^{\nu}_{(i)} \, w^{\rho}_{(0)} \, w^{\sigma}_{(j)}. \label{eq_e_15}
\end{equation}
The symmetric traceless tidal tensor represents the first quadrupolar contribution of the tidal potential
\begin{equation}
\Phi_{\mathrm{t}} = - \frac{1}{2} z^{i} z^{j} \, C_{ij} + O \left( \vert \mathbf{z} \vert^{3} \right). \label{eq_e_16}
\end{equation}
The higher-order contributions relating the deviation of the star's orbit from a strict geodesic can be neglected considering the small size of the star.

It is possible to directly connect the tetrad to the coordinates of the star's centre of mass along its geodesic \citep{mar1983}. It comes
\begin{eqnarray}
w^{\mu}_{(0)} & = & \dot{t} \, \delta^{\mu}_{0} + \dot{r} \, \delta^{\mu}_{1} + \dot{\phi} \, \delta^{\mu}_{2}, \label{eq_e_17} \\
w^{\mu}_{(1)} & = & \tilde{w}^{\mu}_{(1)} \cos \psi - \tilde{w}^{\mu}_{(2)} \sin \psi, \label{eq_e_18} \\
w^{\mu}_{(2)} & = & \tilde{w}^{\mu}_{(1)} \sin \psi + \tilde{w}^{\mu}_{(2)} \cos \psi, \label{eq_e_19} \\
w^{\mu}_{(3)} & = & \frac{1}{r} \, \delta^{\mu}_{3}, \label{eq_e_20}
\end{eqnarray}
where
\begin{eqnarray}
\tilde{w}^{\mu}_{(1)} & = & \frac{r \, \dot{r}}{\left(1-\frac{2 M_{\bullet}}{r}\right) \left(r^{2}+L^{2}\right)^{1/2}} \, \delta^{\mu}_{0}
+ \frac{E r}{\left(r^{2}+L^{2}\right)^{1/2}} \, \delta^{\mu}_{1}, \label{eq_e_21} \\
\tilde{w}^{\mu}_{(2)} & = & \frac{E L}{\left(1-\frac{2 M_{\bullet}}{r}\right) \left(r^{2}+L^{2}\right)^{1/2}} \, \delta^{\mu}_{0} 
+ \frac{L \, \dot{r}}{\left(r^{2}+L^{2}\right)^{1/2}} \, \delta^{\mu}_{1} \nonumber \\
&& \hspace{3.5cm} + \frac{\left(r^{2}+L^{2}\right)^{1/2}}{r^{2}} \, \delta^{\mu}_{2}, \label{eq_e_22}
\end{eqnarray}
with $\delta^{\mu}_{\nu}$ the Kronecker's symbol, and where the $\psi$ variable satisfies
\begin{equation}
\dot{\psi} = \frac{E L}{r^{2}+L^{2}}. \label{eq_e_23}
\end{equation}
From (\ref{eq_e_17})-(\ref{eq_e_22}) and the Riemann curvature tensor, the non-zero components of the tidal tensor (\ref{eq_e_15}) for a Schwarzschild space-time take the explicit form
\begin{eqnarray}
C_{11} & = & - \frac{M_{\bullet}}{r^{3}} \left( 1 - 3 \frac{r^{2} + L^{2}}{r^{2}} \cos^{2} \psi \right), \label{eq_e_24} \\
C_{22} & = & - \frac{M_{\bullet}}{r^{3}} \left( 1 - 3 \frac{r^{2} + L^{2}}{r^{2}} \sin^{2} \psi \right), \label{eq_e_25} \\
C_{33} & = & - \frac{M_{\bullet}}{r^{3}} \left( 1 + 3 \frac{L^{2}}{r^{2}} \right), 
\label{eq_e_26} \\
C_{12} & = & - \frac{M_{\bullet}}{r^{3}} \left( - 3 \frac{r^{2} + L^{2}}{r^{2}} \cos \psi \sin \psi \right), \label{eq_e_27} \\
C_{21} & = & C_{12}.
\label{eq_e_28}
\end{eqnarray}

We assume that the star draws a parabolic-type orbit around the black hole for which $E=1$ and $L \ge 4 M_{\bullet}$.
When the last inequality arises, the black hole actually generates the gravitational barrier which can reject the incoming star to infinity \citep[e.g.][]{cha1983}.
From the cancellation of (\ref{eq_e_03}), the periastron of the orbit is located at
\begin{equation}
r_{\mathrm{p}} = \frac{L^{2}}{4 M_{\bullet}} \left( 1 + \sqrt{ 1 - 16 \left( \frac{M_{\bullet}}{L} \right)^{2} } \right). \label{eq_e_06}
\end{equation}
The critical value $L=4 M_{\bullet}$ dividing plunge and escape orbits corresponds to the minimum periastron
\begin{eqnarray}
r_{\mathrm{p \, m}} & = & 4 \, M_{\bullet} \label{eq_e_07} \\
                    & = & 2 \, r_{g}, \label{eq_e_08}
\end{eqnarray}
where $r_{\mathrm{g}} = 2 \, M_{\bullet}$ is the gravitational radius of the black hole.
From (\ref{eq_e_06}), the specific orbital angular momentum defining the parabolic orbit is thus given by
\begin{equation}
L = r_{\mathrm{p}} \left( \frac{r_{\mathrm{g}}}{r_{\mathrm{p}}-r_{\mathrm{g}}} \right)^{1/2}, \label{eq_e_08_01}
\end{equation}
where the periastron is set through the value of the penetration factor 
\begin{equation}
\beta = \frac{r_{\mathrm{t}}}{r_{\mathrm{p}}} \label{eq_e_08_02}
\end{equation}
of the orbit within the characteristic disruption tidal radius
\begin{equation}
r_{\mathrm{t}} = R_{\star} \left( \frac{M_{\bullet}}{M_{\star}} \right)^{1/3}, \label{eq_e_08_03}
\end{equation}
with $M_{\star}$ and $R_{\star}$ respectively the mass and radius of the star.

To solve the orbital equations (\ref{eq_e_02})-(\ref{eq_e_05}) and (\ref{eq_e_23}) of parabolic motion with $E=1$, we rewrite them as \citep{fro1998}
\begin{eqnarray}
\dot{\tilde{t}}         & = & \left( 1 - \frac{\mu^{2}}{\tilde{r}} \right)^{-1}, \label{eq_e_29} \\
\dot{\tilde{r}}^{2} & = & \frac{1}{\tilde{r}} - \frac{1}{\tilde{r}^{2}} + \frac{\mu^{2}}{\tilde{r}^{3}}, \label{eq_e_30} \\
\dot{\theta}            & = & 0, \label{eq_e_31} \\
\dot{\phi}              & = & \frac{1}{\tilde{r}^{2}}, \label{eq_e_32} \\
\dot{\psi}              & = & \frac{1}{\tilde{r}^{2} + \mu^{2}}, \label{eq_e_33}
\end{eqnarray}
defining
\begin{eqnarray}
\widetilde{L}  & = & \frac{L}{r_{\mathrm{g}}}, \label{eq_e_34} \\
R_{\mathrm{p}} & = & \widetilde{L}^{2} \, r_{\mathrm{g}}, \label{eq_e_35} \\
t_{\mathrm{p}} & = & \widetilde{L}^{3} \, r_{\mathrm{g}}, \label{eq_e_36} \\
\mu            & = & \left( \frac{r_{\mathrm{g}}}{R_{\mathrm{p}}} \right)^{1/2} \label{eq_e_37} \\
               & = & \widetilde{L}^{-1}, \nonumber
\end{eqnarray}
and switching to dimensionless variables
\begin{eqnarray}
\tilde{\tau} & = & \frac{\tau}{t_{\mathrm{p}}}, \label{eq_e_38} \\
\tilde{t}    & = & \frac{t}{t_{\mathrm{p}}}, \label{eq_e_39} \\
\tilde{r}    & = & \frac{r}{R_{\mathrm{p}}}, \label{eq_e_40}
\end{eqnarray}
with $\left(\, \dot{} \, \right)$ the derivative with respect to $\tilde{\tau}$. 
By convention $\tilde{\tau}$, $\phi$, and $\psi$ are set null at the periastron of the orbit, and are negative and positive respectively for earlier and later positions.
For $\tilde{\tau} \ge 0$, the temporal integration of (\ref{eq_e_30}), (\ref{eq_e_32}), and (\ref{eq_e_33}) leads to 
\begin{eqnarray}
\tilde{\tau} & = & \frac{2}{3} \, 
\Bigg(
\frac{\sqrt{\tilde{r}} \sqrt{\tilde{r}^{2} - \tilde{r} + \mu^{2}} \, (\tilde{r} + 2 - \mu)}{\tilde{r} - \mu} \nonumber \\
&& \hspace{0.2cm} + \frac{1 - \mu^{2}}{\sqrt{1 + 2 \mu}} \, F(\varphi \setminus \alpha) - \sqrt{1 + 2 \mu} \, E(\varphi \setminus \alpha)
\Bigg), \label{eq_e_45} \\
\phi & = & \frac{2}{\sqrt{1 + 2 \mu}} \, F(\varphi \setminus \alpha), \label{eq_e_46} \\
\psi & = & \arctan \sqrt{\tilde{r} - 1 + \frac{\mu^{2}}{\tilde{r}}} + \frac{1}{\sqrt{1 + 2 \mu}} F(\varphi \setminus \alpha), \label{eq_e_47}
\end{eqnarray}
introducing the elliptical integrals of the first and second kind, respectively
\begin{eqnarray}
F(\varphi \setminus \alpha) & = & \int_{0}^{\varphi} \, (1 - \sin^{2}\alpha \, \sin^{2}\beta)^{-1/2} \, d \beta, \label{eq_e_41} \\
E(\varphi \setminus \alpha) & = & \int_{0}^{\varphi} \, (1 - \sin^{2}\alpha \, \sin^{2}\beta)^{1/2} \, d \beta, \label{eq_e_42}
\end{eqnarray}
with 
\begin{eqnarray}
\varphi         & = & \arcsin \frac{\sqrt{\tilde{r}^{2} - \tilde{r} + \mu^{2}}}{\tilde{r} - \mu}, \label{eq_e_43} \\ 
\sin^{2} \alpha & = & \frac{4 \mu}{1 + 2 \mu}. \label{eq_e_44}
\end{eqnarray}
The equivalent expressions for $\tilde{\tau} < 0$ deduce from the previous ones taking into account that $\tilde{r}$ is an even function of $\tilde{\tau}$, whereas $\phi$ and $\psi$ are odd functions of $\tilde{\tau}$.
\subsection{Hydrodynamical equations}
Except for the definition of the tidal field, the hydrodynamical model closely follows the assumptions made in \citetalias{bra2008}.
Since the size of the star is excessively small relative to the curvature radius of the external space-time, it is quite possible to keep a Newtonian description of its internal motions.
The hydrodynamics is described in the comoving reference frame that is parallely propagated along the geodesic of the star's centre of mass.
We still restrict to simulate the one-dimensional motion of the stellar matter along the orthogonal direction to the orbital plane when the star is moving within the tidal radius.
As emphasized in \citetalias{bra2008}, such an approximation (compared to full three-dimensional hydrodynamics) is well justified when the star evolution is calculated only during the free-fall and bounce-expansion phases, since then the vertical motion is fully decoupled from the induced motions in the orbital plane, and the latter are quite negligible compared to the dynamics in the vertical direction.

Setting a vertical $z$ axis from the centre to the poles of the star, the Euler's equations write
\begin{equation}
\partial_{\tau} \, \vec{U} + \partial_{z} \, \vec{F}(\vec{U}) = \vec{S}(\vec{U},z,\tau), \label{eq_e_48}
\end{equation}
defining
\begin{eqnarray}
\vec{U}                    & = & \left[ \rho,   \rho v,         \rho e           \right]^{T}, \label{eq_e_49} \\
\vec{F} ( \vec{U} )        & = & \left[ \rho v, P + \rho v^{2}, ( P + \rho e ) v \right]^{T}, \label{eq_e_50} \\
\vec{S} ( \vec{U},z,\tau ) & = & \left[ 0,      \rho g,         \rho v g         \right]^{T}, \label{eq_e_51}
\end{eqnarray}
respectively the conserved variables, the fluxes, and the sources, where $\rho$ is the density, $P$ the pressure, $v$ the velocity, $e$ the specific total energy, and $g$ the tidal acceleration.

The specific total energy 
\begin{equation}
e=\varepsilon + \frac{v^{2}}{2} \label{eq_e_52}
\end{equation}
includes the specific internal energy $\varepsilon$ and the specific kinetic energy.
The stellar matter is considered as an ideal gas of constant adiabatic index $\gamma=5/3$ for which 
\begin{equation}
P = (\gamma - 1) \rho \varepsilon. \label{eq_e_53}
\end{equation}
The corresponding temperature satisfies
\begin{equation}
T = \frac{\gamma -1}{\mathcal{R}} \varepsilon, \label{eq_e_54}
\end{equation} 
with $\mathcal{R}$ the specific gas constant set for the solar composition. 
The speed of sound is locally defined by
\begin{equation}
a = \sqrt{(\gamma-1) \gamma \varepsilon}. \label{eq_e_55}
\end{equation} 

The general three-dimensional expression of the tidal acceleration is given by (\ref{eq_e_14}) with (\ref{eq_e_24})-(\ref{eq_e_28}).
However, for the one-dimensional model, we only need to take into account the third component which is responsible for the compressive motion orthogonal to the orbital plane.
It simply writes 
\begin{equation}
g(z,\tau) = - \frac{M_{\bullet}}{r^{3}} \left( 1 + 3 \frac{L^{2}}{r^{2}} \right) \, z, \label{eq_e_56}
\end{equation}
with the specific orbital angular momentum of the star defined by (\ref{eq_e_08_01}) and the radial coordinate deduced from the resolution of (\ref{eq_e_45}).

Finally, the conservation form of the Euler's equations (\ref{eq_e_48})-(\ref{eq_e_51}) represents a hyperbolic system of conservation laws.
It is solved by the Godunov-type shock-capturing source-splitting algorithm detailed in \citetalias{bra2008}.
%
%
%
%
\section{Results}
Identically to the Newtonian calculations, we are interested in typical encounters between a black hole of mass $M_{\bullet}=10^{6} \, M_{\odot}$, and a main-sequence star of mass $M_{\star}=M_{\odot}$ and radius $R_{\star}=R_{\odot}$ taken as an initial polytrope of polytropic index $3/2$.
The initial central density, pressure, and temperature are respectively noted $\rho_{\star}$, $P_{\star}$, and $T_{\star}$. 
We consider cases where the star deeply plunges with $\beta>1$ within the tidal radius (\ref{eq_e_08_03}) to simulate its resulting orthogonal compression.
The existence of the minimum periastron (\ref{eq_e_07}) however imposes a maximum penetration factor $\approx 11$ (above this value the star enters the black hole).

Relativistic modifications on the stellar orbit and on the tidal field significantly appear when the star approaches the black hole's gravitational radius. Due to relativistic precession, the parabolic-type orbit must eventually  intersect once in the Schwarzschild space-time, and depending on the position of the crossing point, the star can be subjected to several successive compressions during its motion within the tidal radius \citep{lum1985a,lag1993a}.
\\
\\
\noindent
\textbf{Crossing orbit outside the tidal radius}

\begin{figure}
\begin{tabular}{cc}
\hspace{-0.25cm}
\includegraphics[width=5cm]{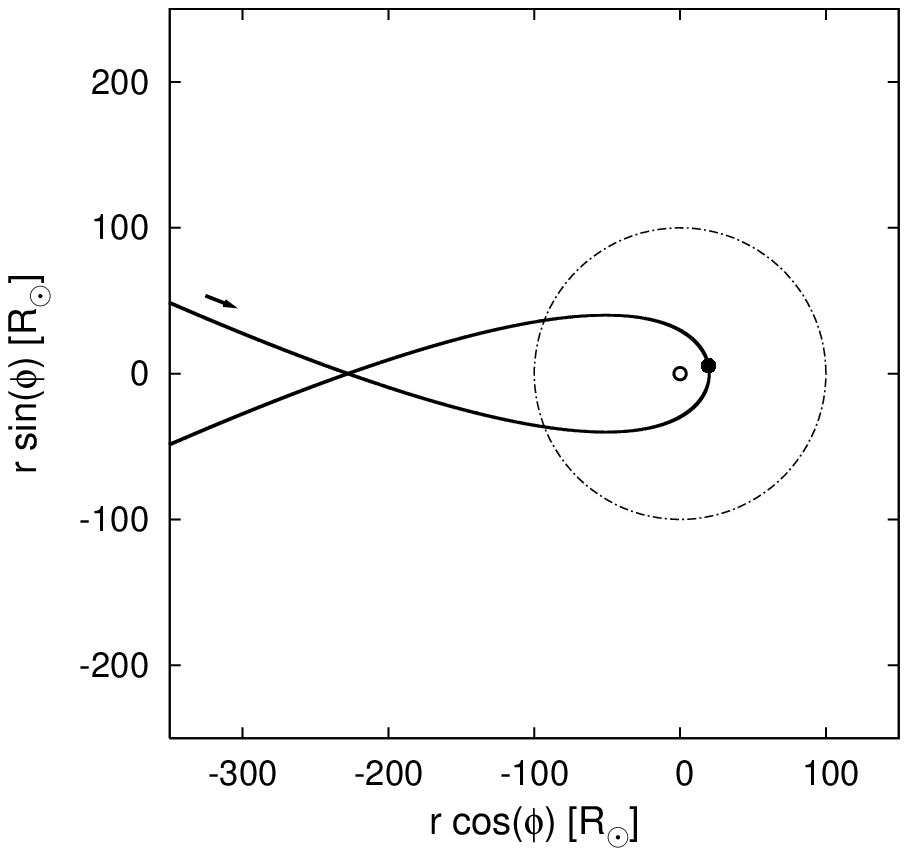} &
\hspace{-1.5cm}
\includegraphics[width=5cm]{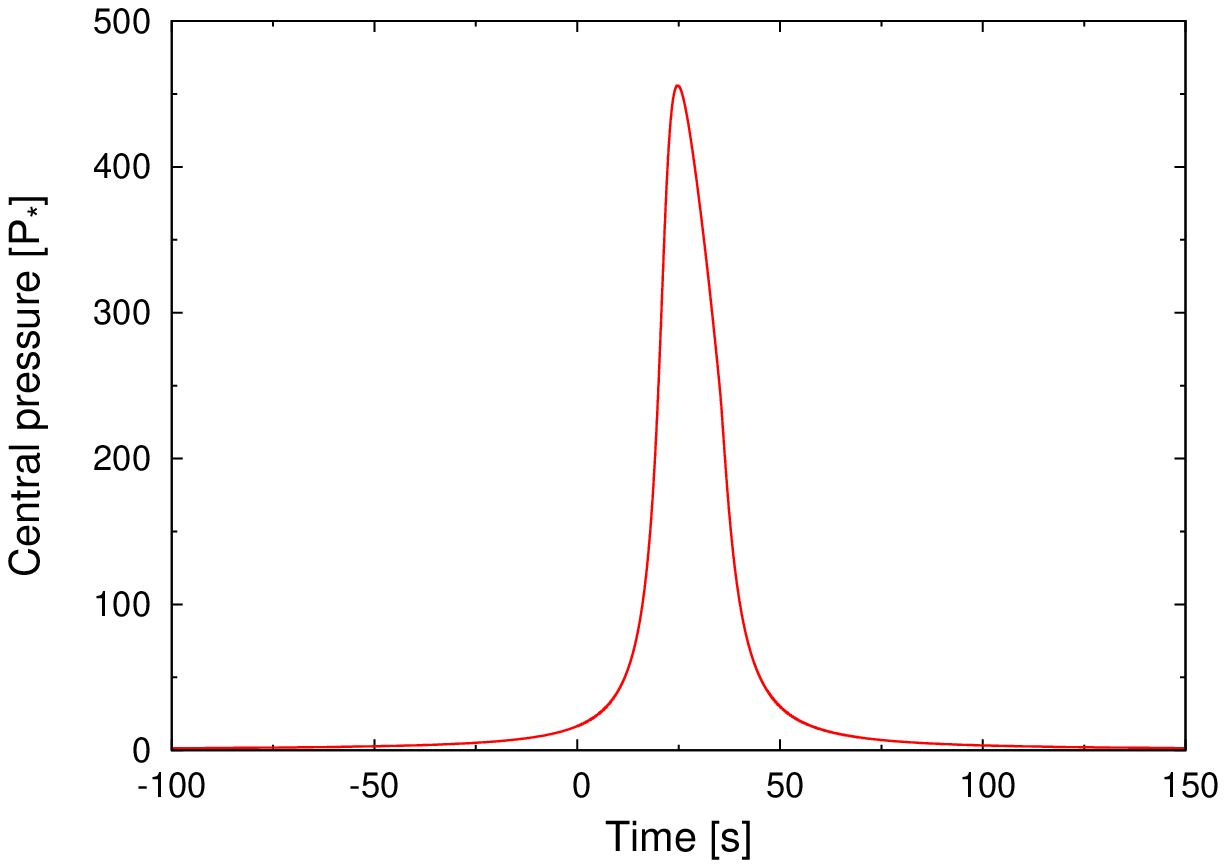}
\end{tabular}
\caption{
Left: Parabolic-type geodesic orbit of a solar-type star around a $10^{6} \, M_{\odot}$ Schwarzschild black hole deeply plunging within the tidal radius with a penetration factor $\beta=5$.
The solid circle represents the black hole's gravitational radius and the dot-dashed circle the tidal radius.
The point on the orbit shows the instant of maximum compression at the centre of the star.
The stellar orbit within the tidal radius does not significantly differ from its Newtonian counterpart, and finally intersect outside the tidal radius so that the tidal field induces one compression.
Right: Evolution of the central pressure as a function of proper time $\tau$.
After the passage of the star through the periastron at $\tau = 0$, the stellar matter suddenly compresses at the centre of the star until the pressure reaches the magnitude needed to counteract the compression's contribution of the black hole's tidal field. 
The central pressure is highest at $\tau \approx 24.66$.
}
\label{fig_g_01}
\end{figure}

\begin{figure}
\begin{tabular}{cc}
\hspace{-0.25cm}
\includegraphics[width=4.5cm]{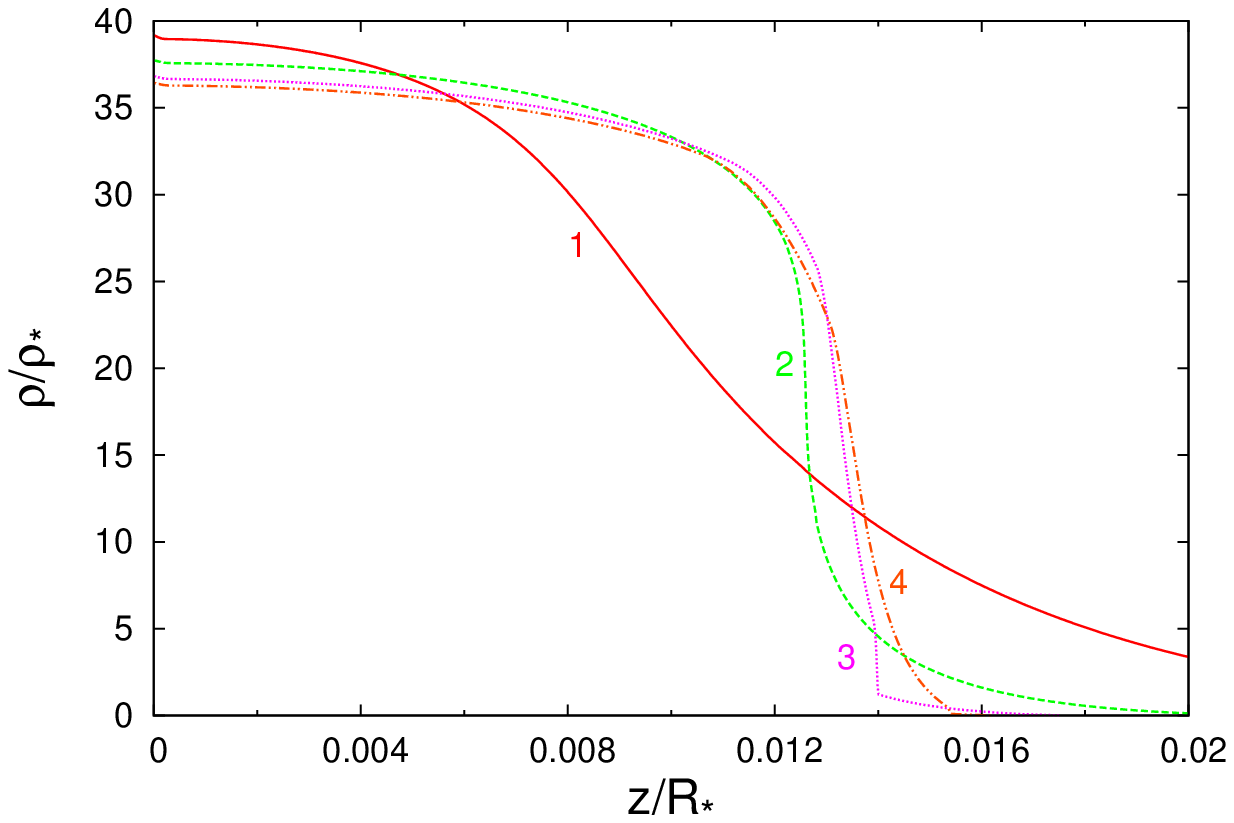} &
\hspace{-0.5cm}
\includegraphics[width=4.5cm]{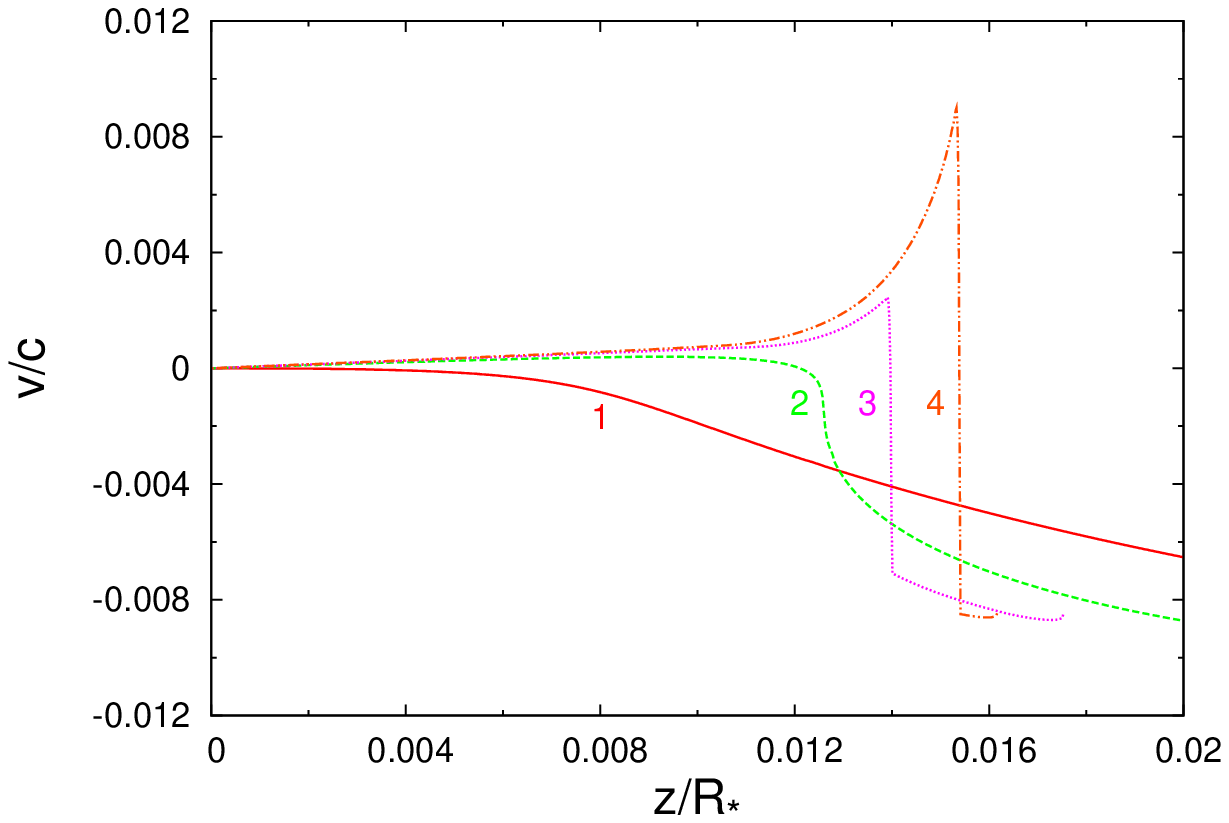} 
\end{tabular}
\caption{
Density (left) and velocity (right) profiles in the positive vertical direction $z$ at different proper times $\tau$ during the bounce-expansion phase for $\beta=5$.
Labels stand for the following values of $\tau$ $[\mathrm{s}]$: (1)~24.66, (2)~27.43, (3)~28.39, (4)~28.75.
The star is at the periastron at $\tau=0$. 
The central compression is maximal at $\tau \approx 24.66$.
The stellar matter expands from the centre whereas it collapses elsewhere.
Both opposite motions produce pressure waves that steepen into a shock wave.
Powered by the expanding motion, the shock wave propagates outwards through the collapsing matter with a velocity $\approx 10^{3} \, \mathrm{km \, s^{-1}}$, and induces a strong compression ratio $\approx 3.5$.
The collapse stops at $\tau \approx 28.80$ when the shock wave reaches the stellar surface $\approx 1.4 \, \mathrm{s}$ after its formation.
}
\label{fig_g_02}
\end{figure}

A first encounter with $\beta=5$ is illustrated on Fig.~\ref{fig_g_01}.
The star is still far away from the gravitational radius for the space-time curvature to remain weak, and for the crossing point to be located outside the tidal radius.
The description of the tidal compression process therefore proceeds as in the Newtonian case in three distinct phases: over-all free fall, central bounce-expansion with shock waves formation, over-all expansion \citepalias[see][Figs. 4, 6, 7]{bra2008}.
The evolution of the hydrodynamical variables during the bounce-expansion phase can be followed in Fig.~\ref{fig_g_02}.
\\
\\
\noindent
\textbf{Crossing orbit inside the tidal radius}

\begin{figure}
\begin{tabular}{cc}
\hspace{-0.25cm}
\includegraphics[width=5cm]{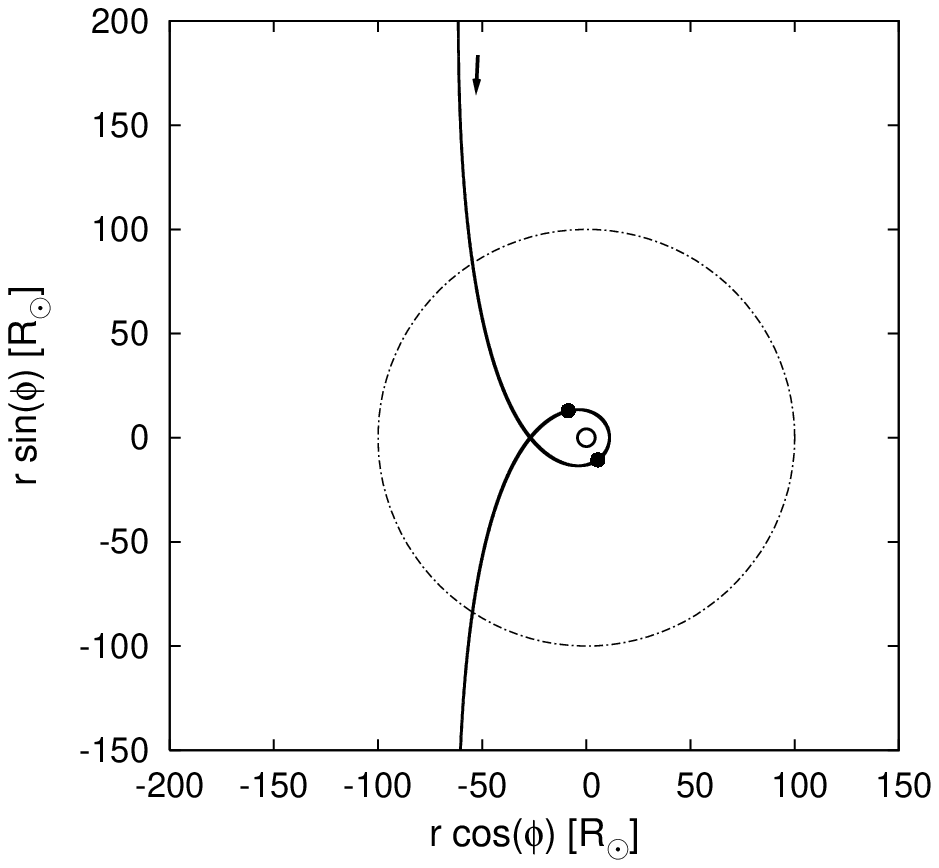} &
\hspace{-1.5cm}
\includegraphics[width=5cm]{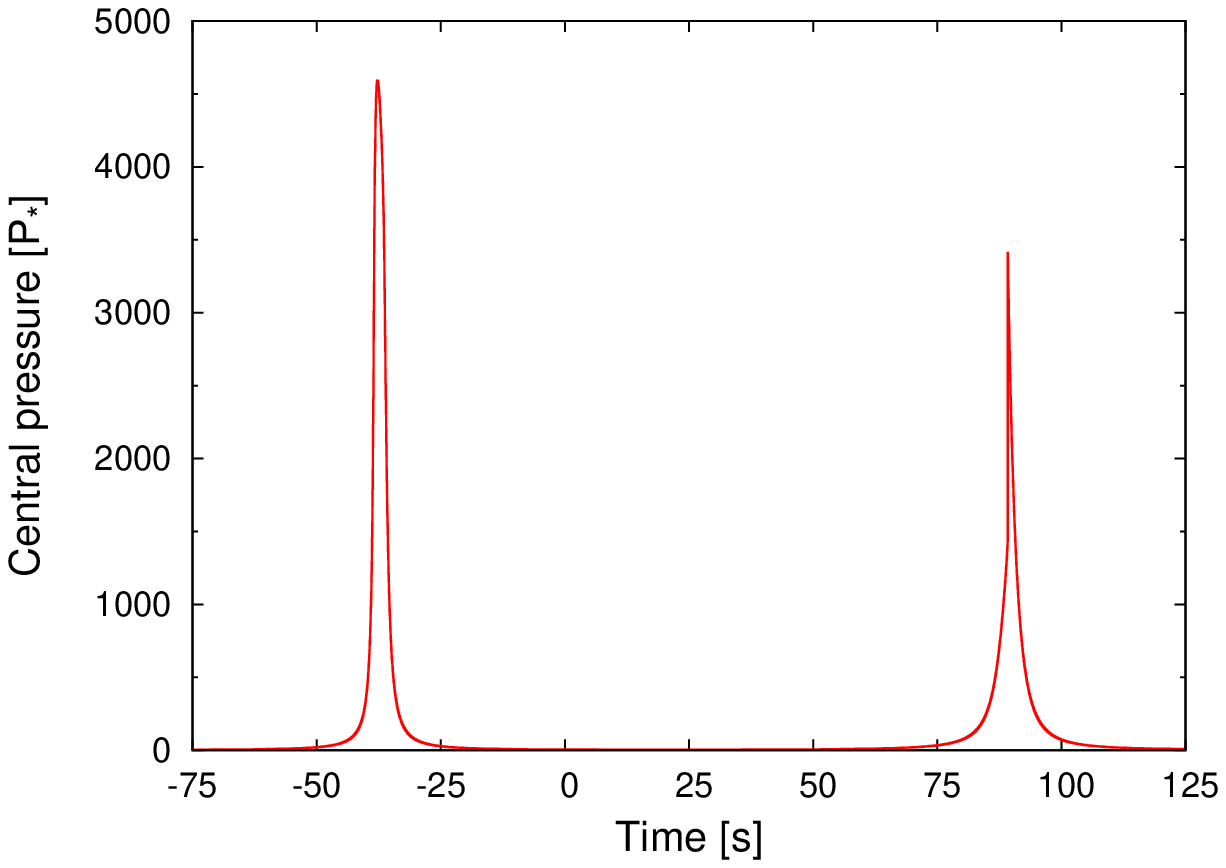}
\end{tabular}
\caption{
Left: Parabolic-type geodesic orbit of a solar-type star around a $10^{6} \, M_{\odot}$ Schwarzschild black hole deeply plunging within the tidal radius with a penetration factor $\beta=9$.
The solid circle represents the black hole's gravitational radius and the dot-dashed circle the tidal radius.
The points on the orbit show the instants of maximum compression at the centre of the star.
The stellar orbit winds up inside the tidal radius so that the tidal field induces two successive compressions.
Right: Evolution of the central pressure as a function of proper time $\tau$.
The central pressure is highest at $\tau \approx -37.68$ before the passage through the periastron, and at $\tau \approx 89.13$ after the passage through the periastron.
}
\label{fig_g_03}
\end{figure}

A second encounter is illustrated on Fig.~\ref{fig_g_03} for a penetration factor $\beta=9$, huge enough for the star to come nearer to the black hole's gravitational radius and experience the space-time curvature.
The stellar orbit winds up around the black hole within the tidal radius, which allows the tidal field to highly compress the star twice before and after the periastron.

\begin{figure}
\begin{tabular}{ccc}
\hspace{-0.3cm}
\includegraphics[width=4.5cm]{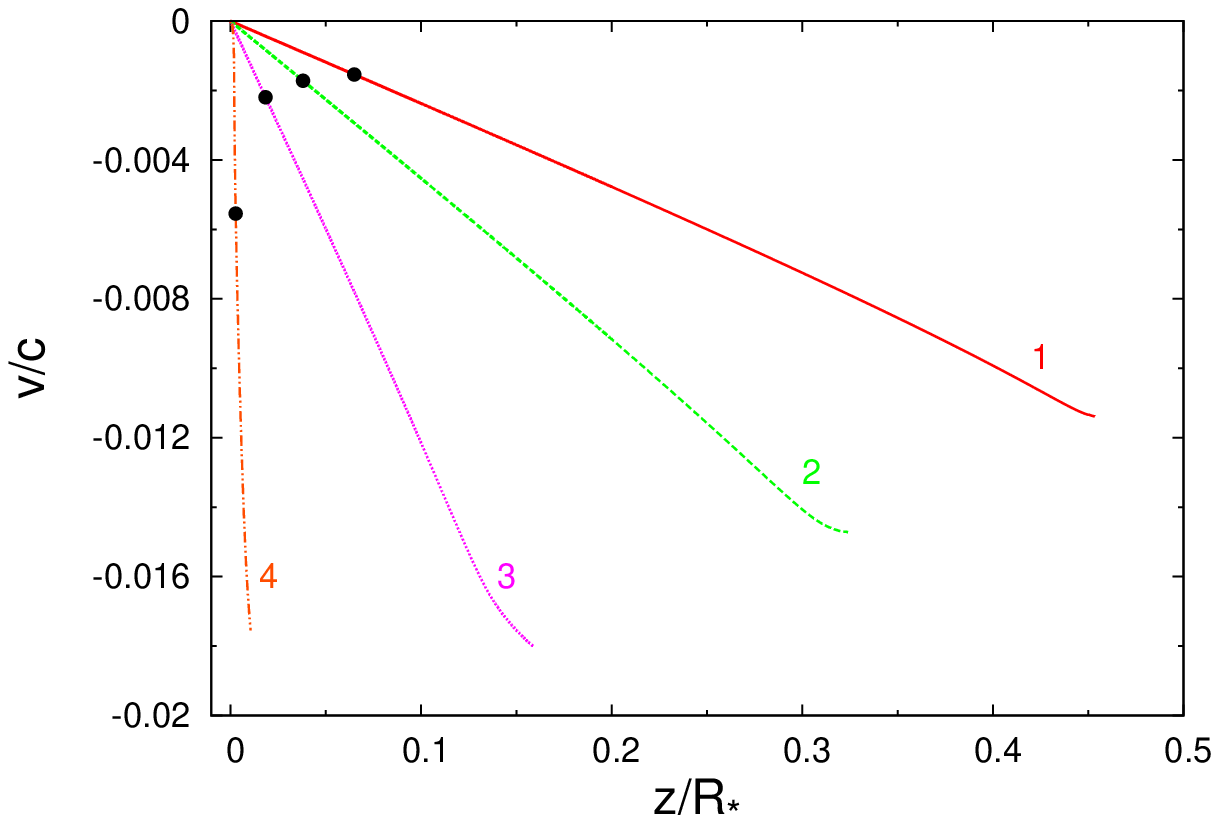} &
\hspace{-0.5cm}
(a) &
\hspace{-0.8cm}
\includegraphics[width=4.5cm]{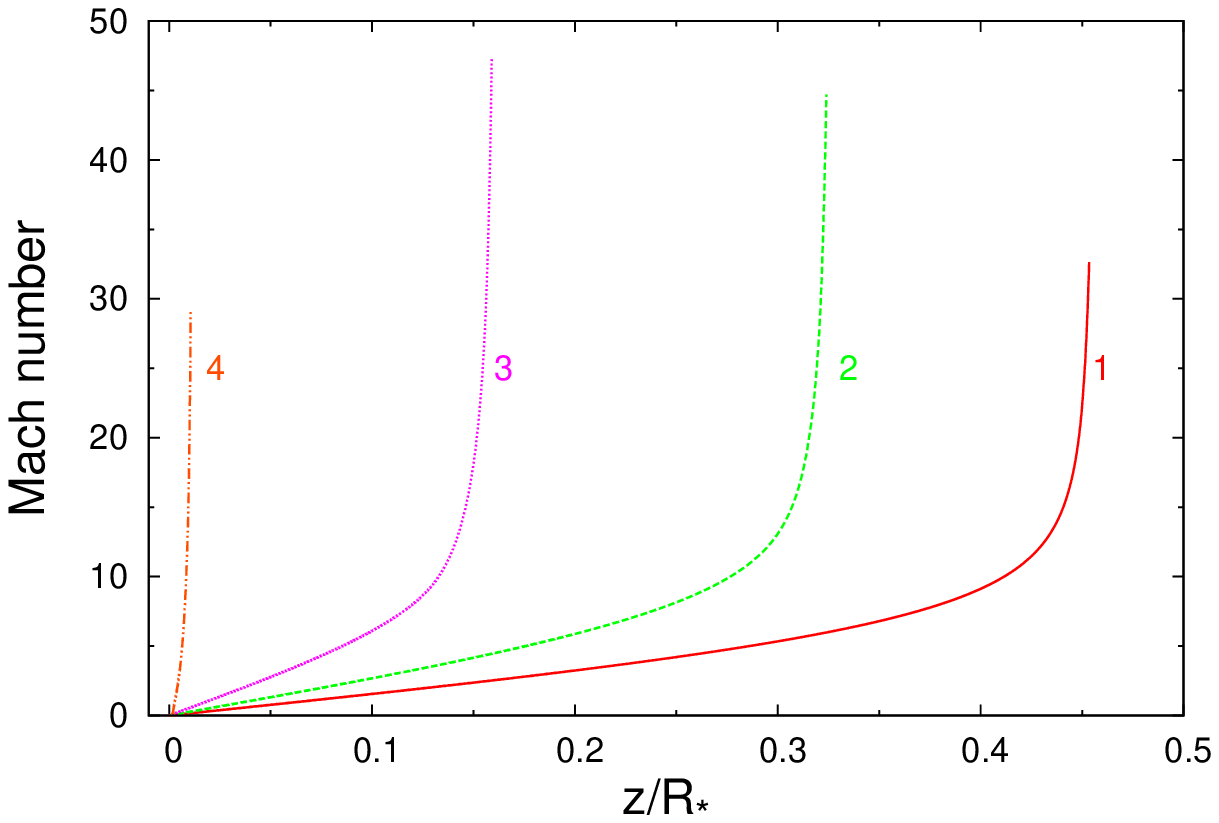} \\
\hspace{-0.3cm}
\includegraphics[width=4.5cm]{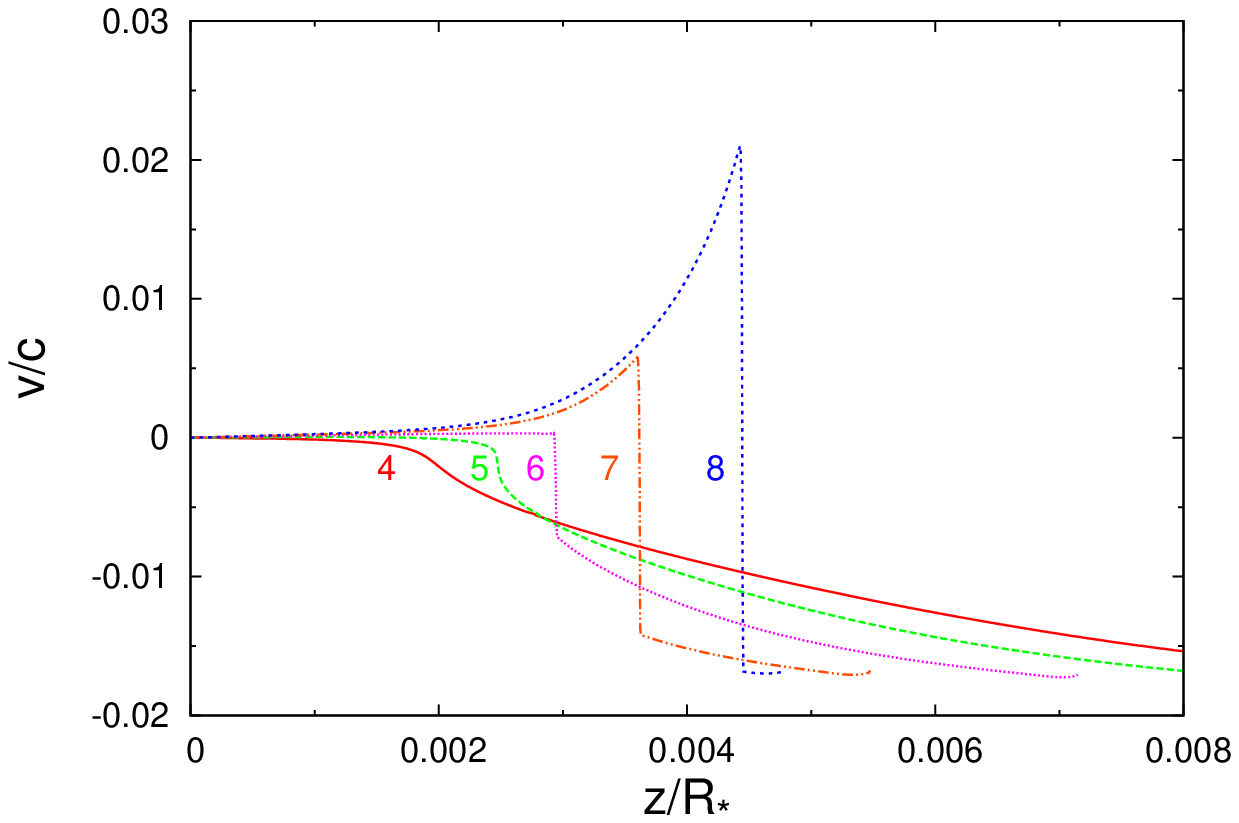} &
\hspace{-0.5cm}
(b) & \\
\hspace{-0.3cm}
\includegraphics[width=4.5cm]{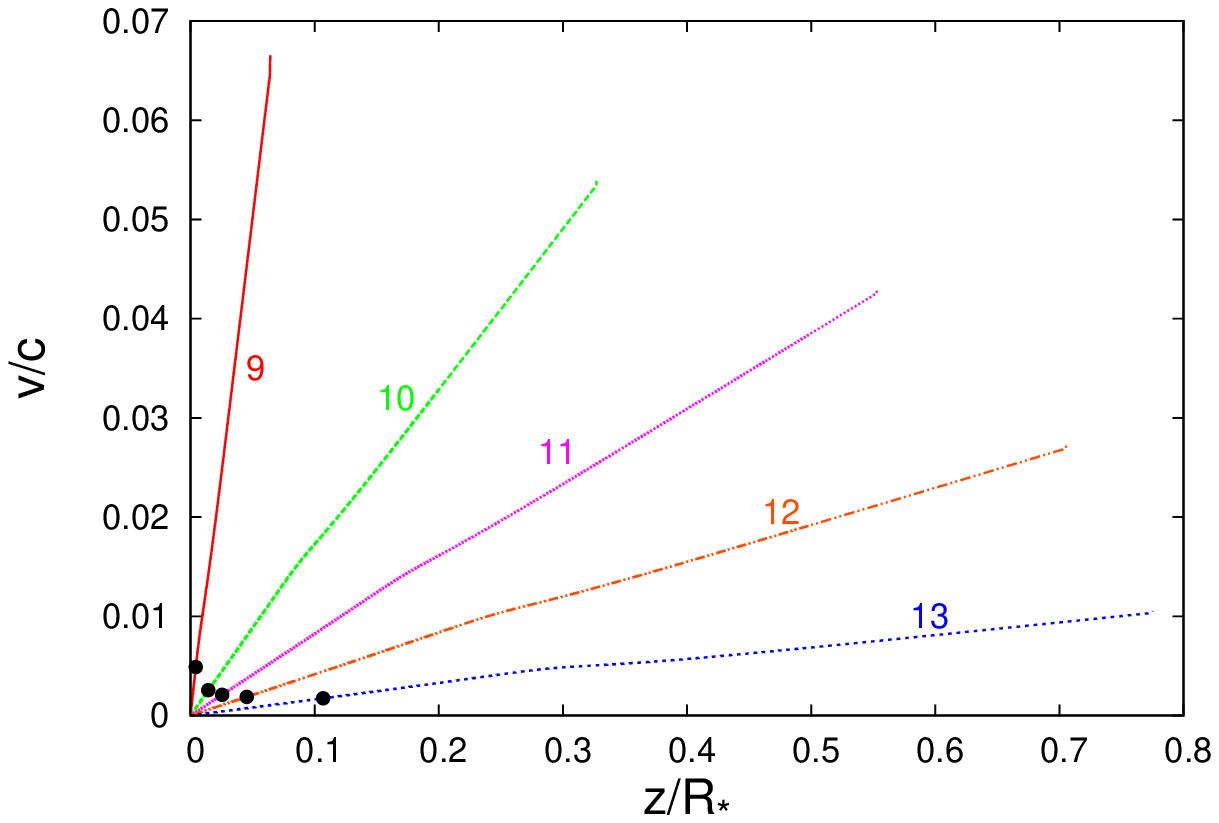} &
\hspace{-0.5cm}
(c) &
\hspace{-0.8cm}
\includegraphics[width=4.5cm]{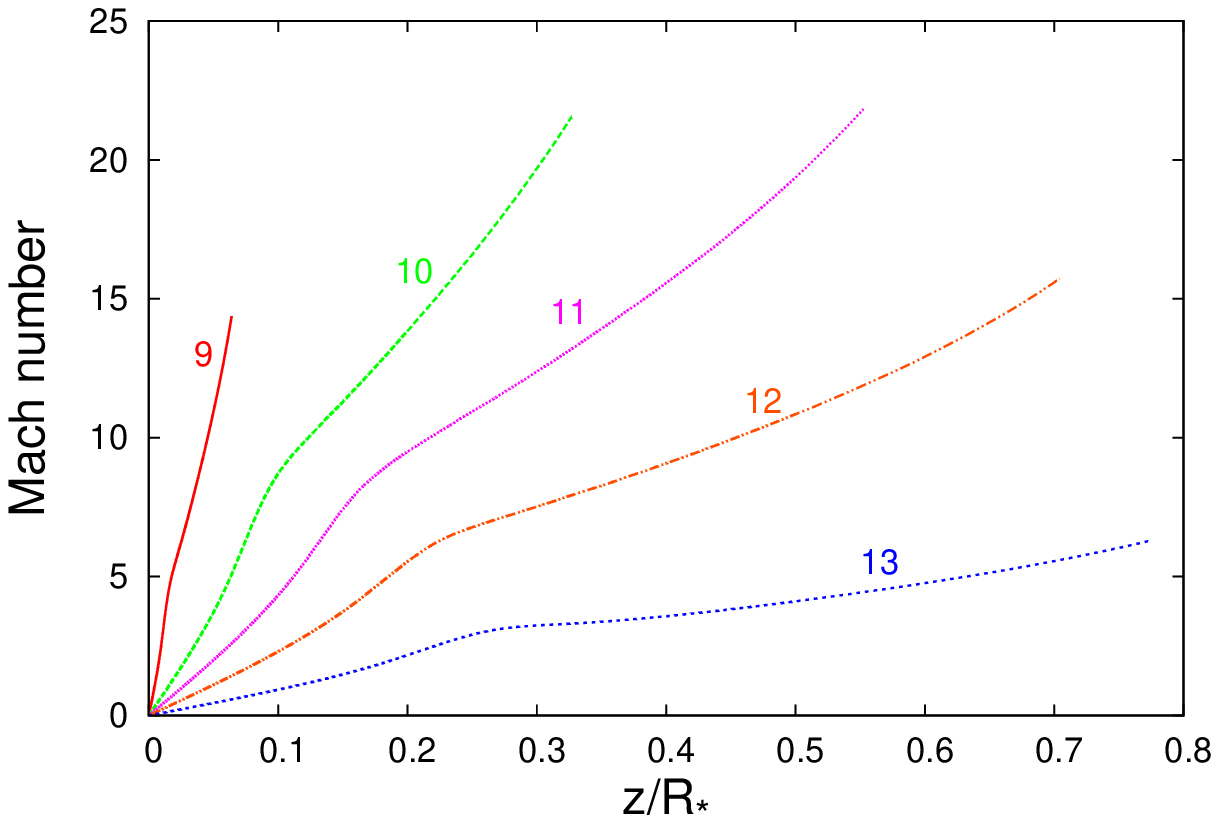}
\end{tabular}
\caption{
Velocity (left) and Mach number (right) profiles in the positive vertical direction $z$ at different proper times $\tau$ during the first compression for $\beta=9$.
The temporal evolution reads from top to bottom.
(a): over-all free fall, (b): central bounce-expansion, (c): over-all expansion.
Labels stand for the following values of $\tau$ $[\mathrm{s}]$: (1)~-102.94, (2)~-79.74, (3)~-56.54, (4)~-37.68, (5)~-37.56, (6)~-37.33, (7)~-37.09, (8)~-37.00, (9)~-34.81, (10)~-23.41, (11)~-11.76, (12)~0.15, (13)~11.60.
The star is at the periastron at $\tau=0$.
The point on the velocity profiles corresponds to the sonic point, the flow being subsonic (resp. supersonic) to the left (resp. right).
The homologous velocity profiles $v(z,\tau) \sim z$ characterize the free-fall and expansion phases in the external gravitational field.
The Mach number reaches remarkably high values.
The stellar surface collapses with a maximum velocity $\approx 5 \times 10^{3} \, \mathrm{km \, s^{-1}}$, and expands with a maximum velocity $\approx 10^{4} \, \mathrm{km \, s^{-1}}$.
The free fall stops from the centre of the star at $\tau \approx -37.68$ when the central pressure reaches its maximum.
The stellar matter then begins to expand while it continues to collapse elsewhere.
Pressure waves steepen into a shock wave $S_{1}$ that propagates outwards with a velocity $\approx 10^{3} \, \mathrm{km \, s^{-1}}$ until it escapes from the medium at $\tau \approx -36.98$.
It induces a strong compression ratio $\approx 3.7$.
The whole stellar matter then continues to expand.
}
\label{fig_g_04}
\end{figure}

\begin{figure}
\resizebox{\hsize}{!}{\includegraphics{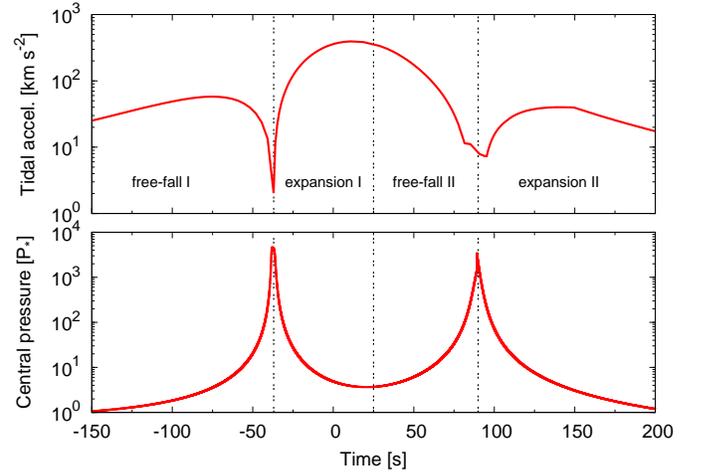}}
\caption{
Evolution of the tidal acceleration at the surface of the star as a function of proper time $\tau$.
The tidal acceleration (\ref{eq_e_56}) is maximal at the stellar surface at each time.
The simultaneous evolution of the central pressure is added for comparison.
The star is at the periastron at $\tau=0$.
The double successive phases of free-fall and expansion are indicated. 
The decreasing of the tidal acceleration during the orbital motion of the star allows the central pressure to suddenly counteract twice the free fall, before and after the periastron passage.
}
\label{fig_g_05}
\end{figure}

The stellar matter enters a \textit{first} supersonic free-fall phase until the central pressure highly increases to make it bounce and stop the compressive motion (Fig.~\ref{fig_g_04}).
As in the previous case, a strong shock wave $S_{1}$ forms outwards during the bounce, and escapes from the stellar medium to give way to an over-all supersonic expansion.
However, the periastron being not yet reached, the tidal field becomes again prevailing relative to the central pressure of the star which decreases as suddenly as it previously increased (Fig.~\ref{fig_g_05}).

\begin{figure}
\begin{tabular}{ccc}
\hspace{-0.3cm}
\includegraphics[width=4.5cm]{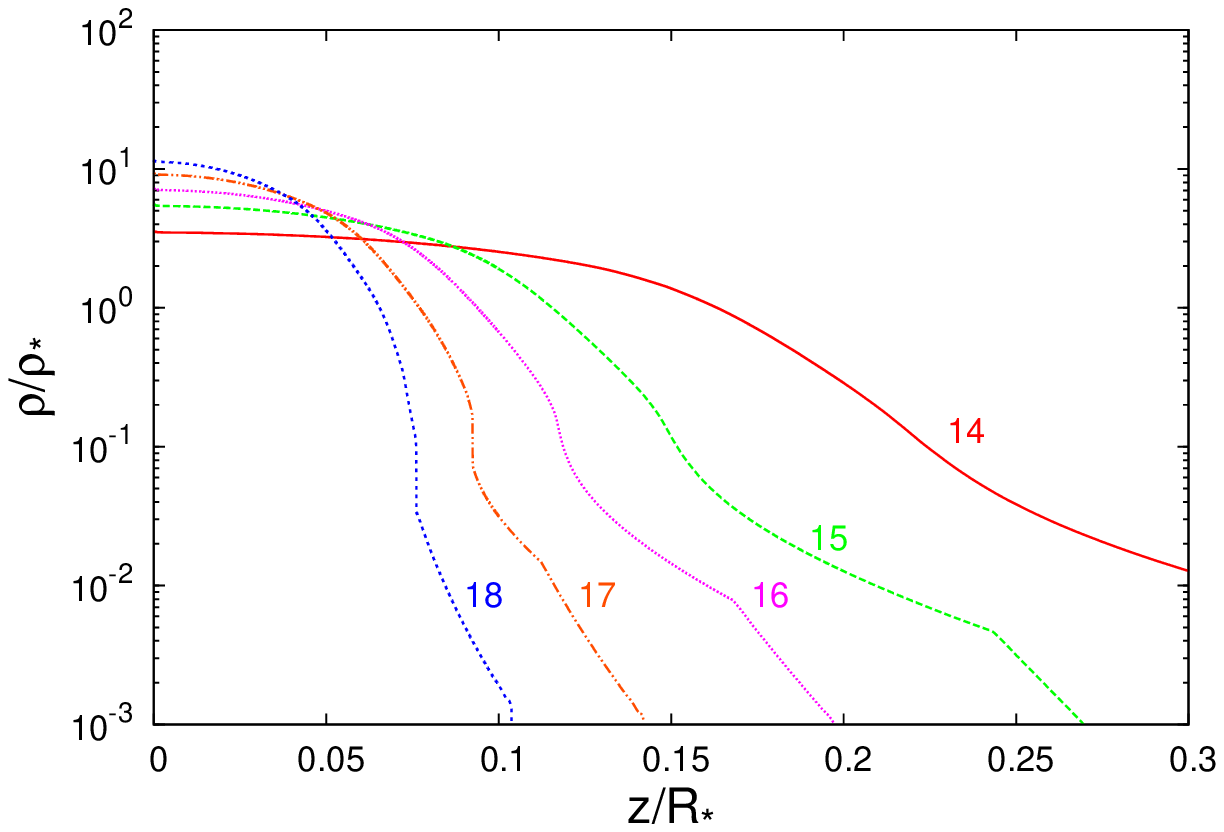} &
\hspace{-0.5cm}
(a) &
\hspace{-0.8cm}
\includegraphics[width=4.5cm]{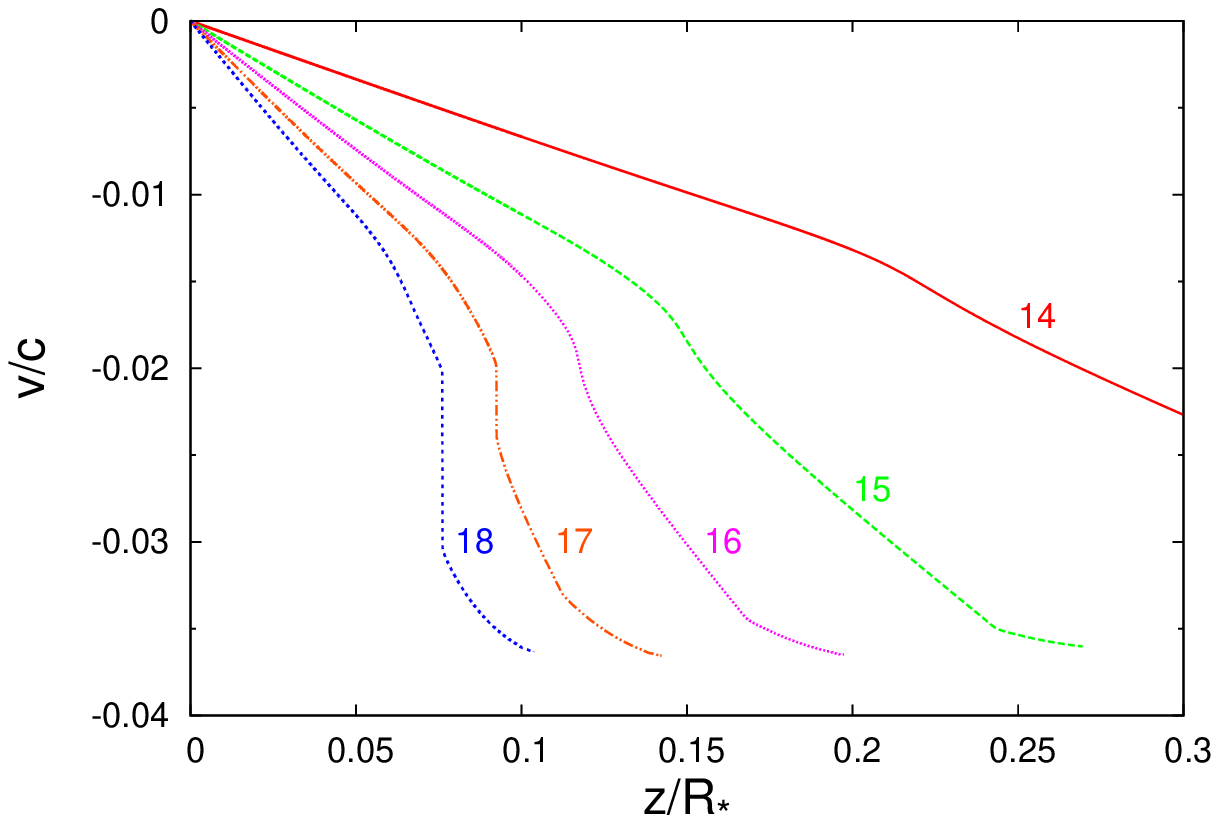} \\
\hspace{-0.3cm}
\includegraphics[width=4.5cm]{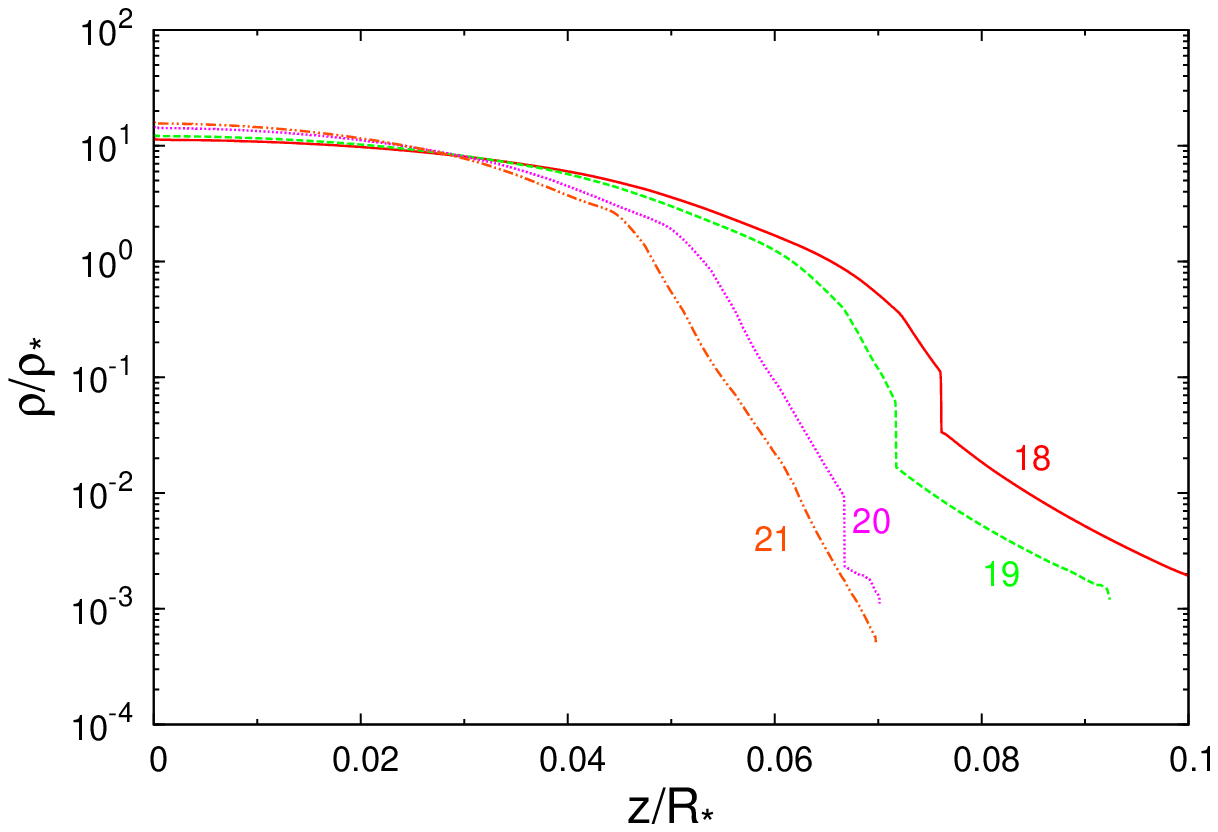} &
\hspace{-0.5cm}
(b) &
\hspace{-0.8cm}
\includegraphics[width=4.5cm]{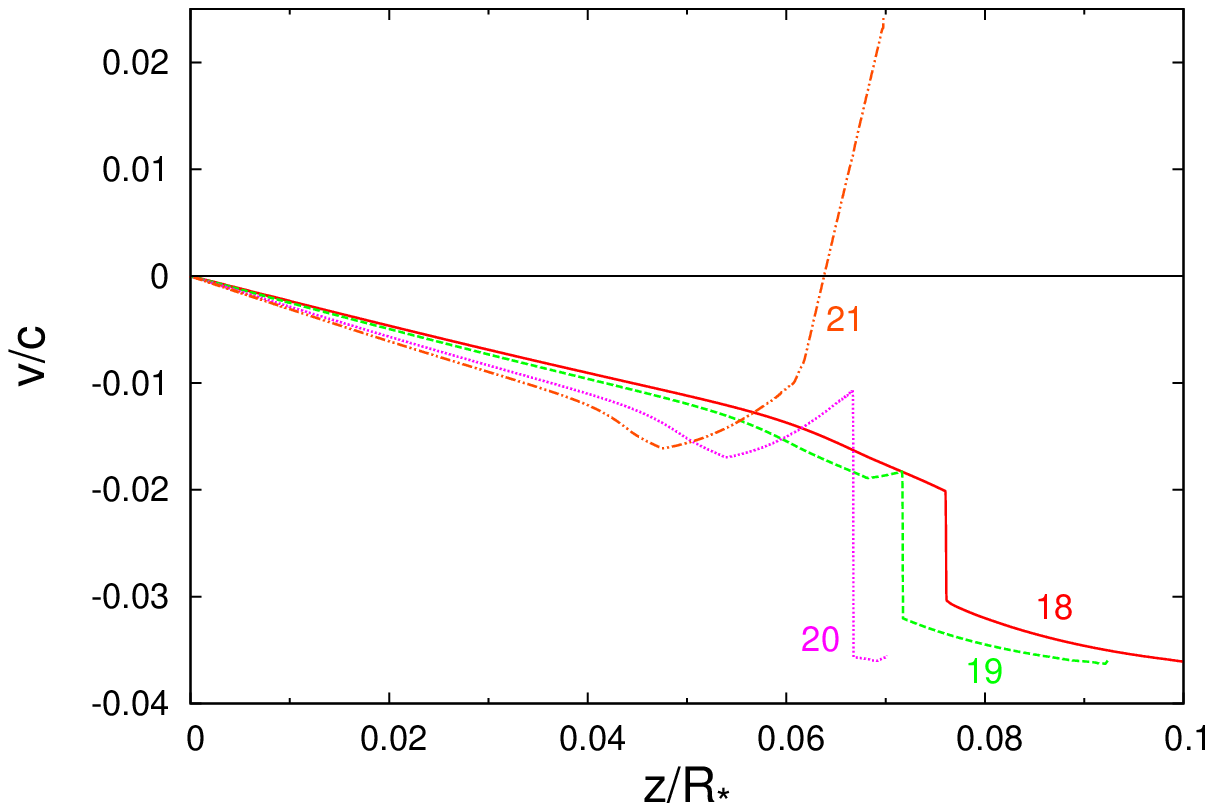} \\
\hspace{-0.3cm}
\includegraphics[width=4.5cm]{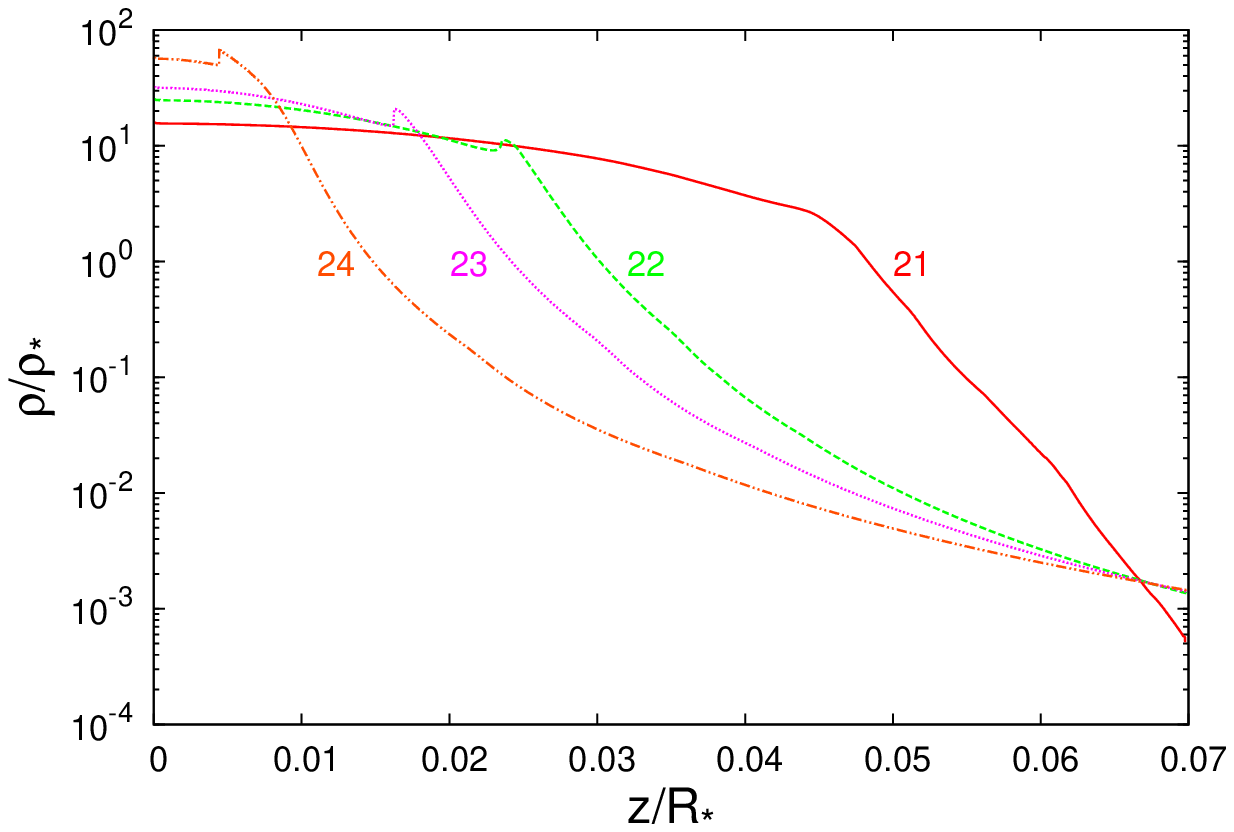} &
\hspace{-0.5cm}
(c) &
\hspace{-0.8cm}
\includegraphics[width=4.5cm]{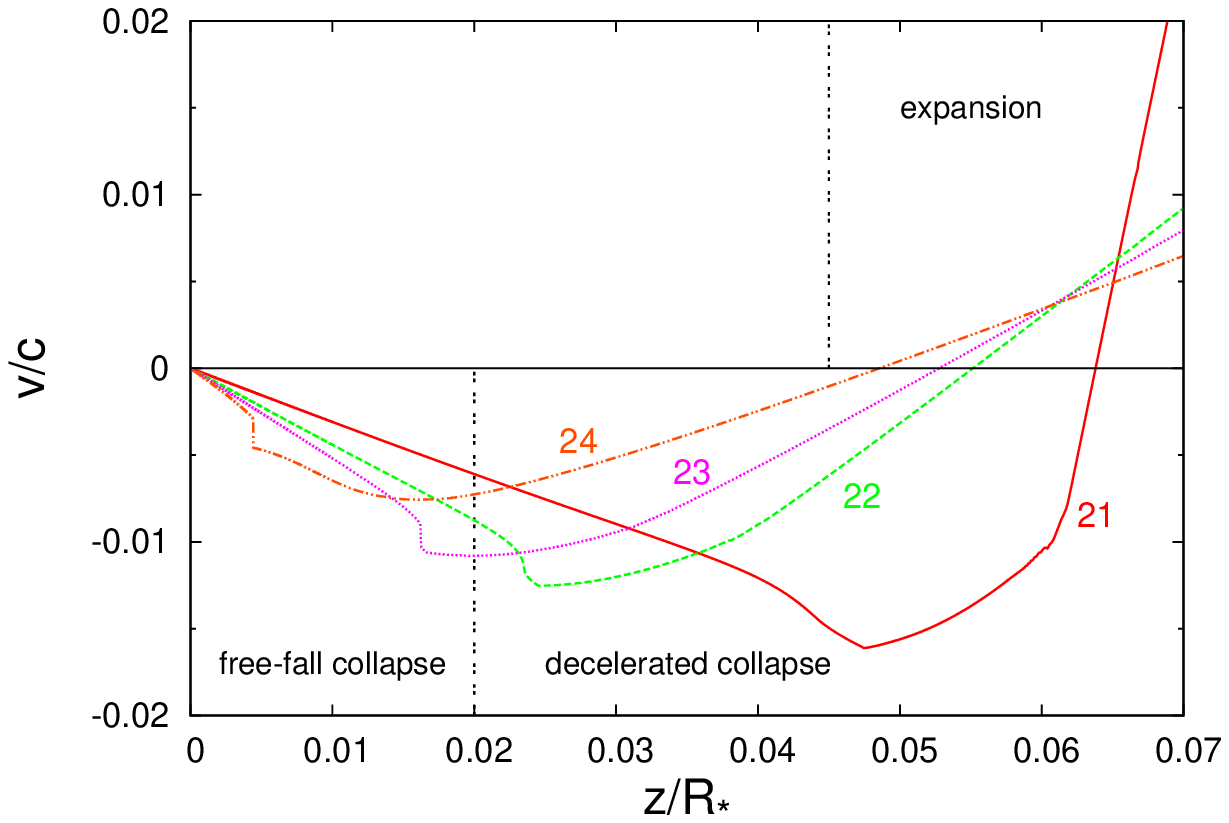}
\end{tabular}
\caption{
Density (left) and velocity (right) profiles in the positive vertical direction $z$ at different proper times $\tau$ during the second free-fall phase for $\beta=9$.
The temporal evolution reads from top to bottom.
Labels stand for the following values of $\tau$ $[\mathrm{s}]$: (14)~56.48, (15)~68.13, (16)~72.80, (17)~76.29, (18)~78.61, (19)~79.31, (20)~80.69, (21)~81.40, (22)~84.31, (23)~85.49, (24)~87.86.
The star is at the periastron at $\tau=0$.
While the stellar matter collapses in free fall, a shock wave $S_{2}$ of strong compression ratio $\approx 3.5$ forms (a).
The shock wave propagates inwards with a velocity $\approx 10^{3} \, \mathrm{km \, s^{-1}}$, whereas the stellar surface collapses with a velocity 10 times superior.
The former is finally caught up by the latter and escapes from the medium without reaching the centre of the star (b).
The velocity profiles (c) show that one part of the stellar matter is slowed down going through the shock wave, while another part stops to collapse near the stellar surface following the ejection of the shock wave, and while in the front the non-affected stellar matter continues to free fall with a homologous velocity profile.
A shock wave $S_{3}$ of weaker compression ratio $\approx 1.5$ forms near the interface of both collapsing parts, and reaches the centre of the star at $\tau \approx 89.13$.
}
\label{fig_g_06}
\end{figure}

\begin{figure}
\resizebox{\hsize}{!}{\includegraphics{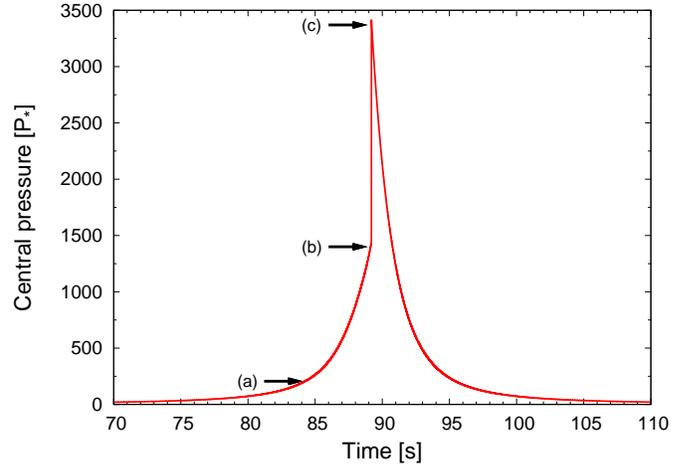}}
\caption{
Evolution of the central pressure as a function of proper time $\tau$ during the second free-fall phase for $\beta=9$ (enlargement of Fig.~\ref{fig_g_03}).
While the stellar matter suddenly compresses for the second time at the centre of the star, a shock wave $S_{3}$ forms at $\tau \approx 84.31$ on both sides of the orbital plane (a).
Both symmetric shock waves propagate inwards until they collide at the centre of the star at $\tau \approx 89.13$ (b), which produces an additional (instantaneous) compression of the central matter (b) $\to$ (c).
The shock waves are then reflected outwards which stops the central compression.
}
\label{fig_g_07}
\end{figure}

The stellar matter is then forced to enter a \textit{second} free-fall phase after the periastron, which occurs differently to the first one (Fig.~\ref{fig_g_06}).
During that phase, a strong shock wave $S_{2}$ forms as was already the case in the Newtonian context for large enough penetration factors \citepalias[see][Figs. 8, 13]{bra2008}.
However, its formation occurs quite further from the centre of the star.
The shock wave propagates slower than the stellar matter collapsing behind it, so that the latter goes through the former which is finally ejected from the medium without reaching the centre of the star.
While the stellar matter continues to collapse, the central pressure begins again to increase (Fig.~\ref{fig_g_07}) following the winding of the stellar orbit, and a third shock wave $S_{3}$ of small amplitude forms.
This shock wave propagates up to the centre where it collides with the symmetric shock wave propagating on the other side of the orbital plane.

\begin{figure}
\begin{tabular}{ccc}
\hspace{-0.3cm}
\includegraphics[width=4.5cm]{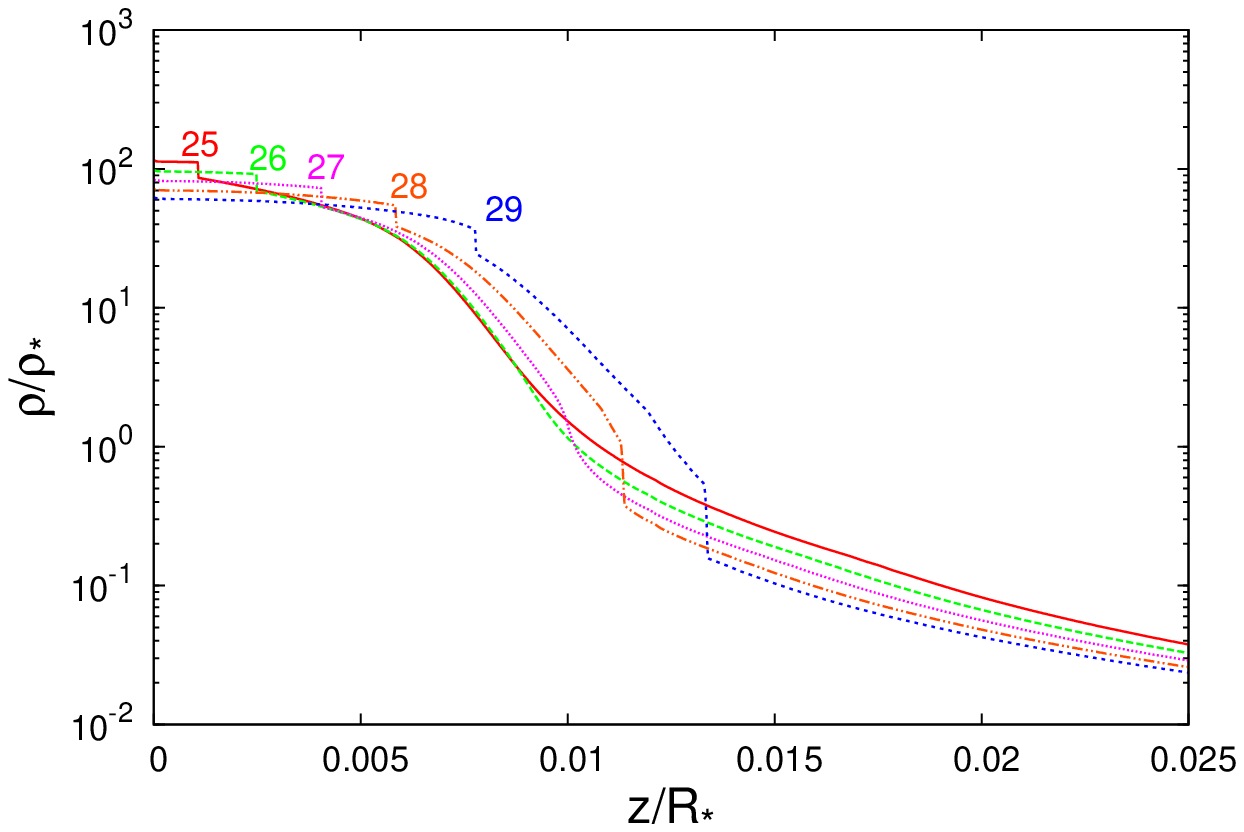} &
\hspace{-0.5cm}
(a) &
\hspace{-0.8cm}
\includegraphics[width=4.5cm]{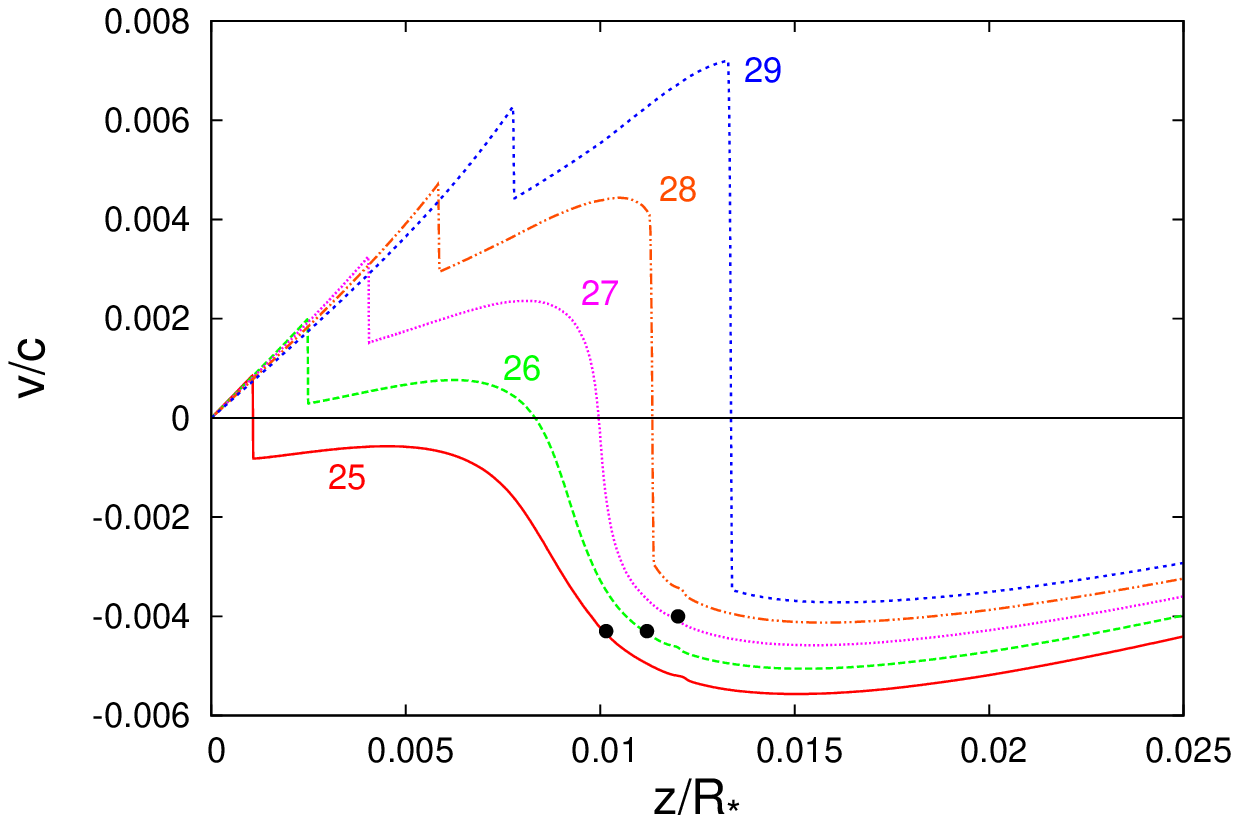} \\
\hspace{-0.3cm}
\includegraphics[width=4.5cm]{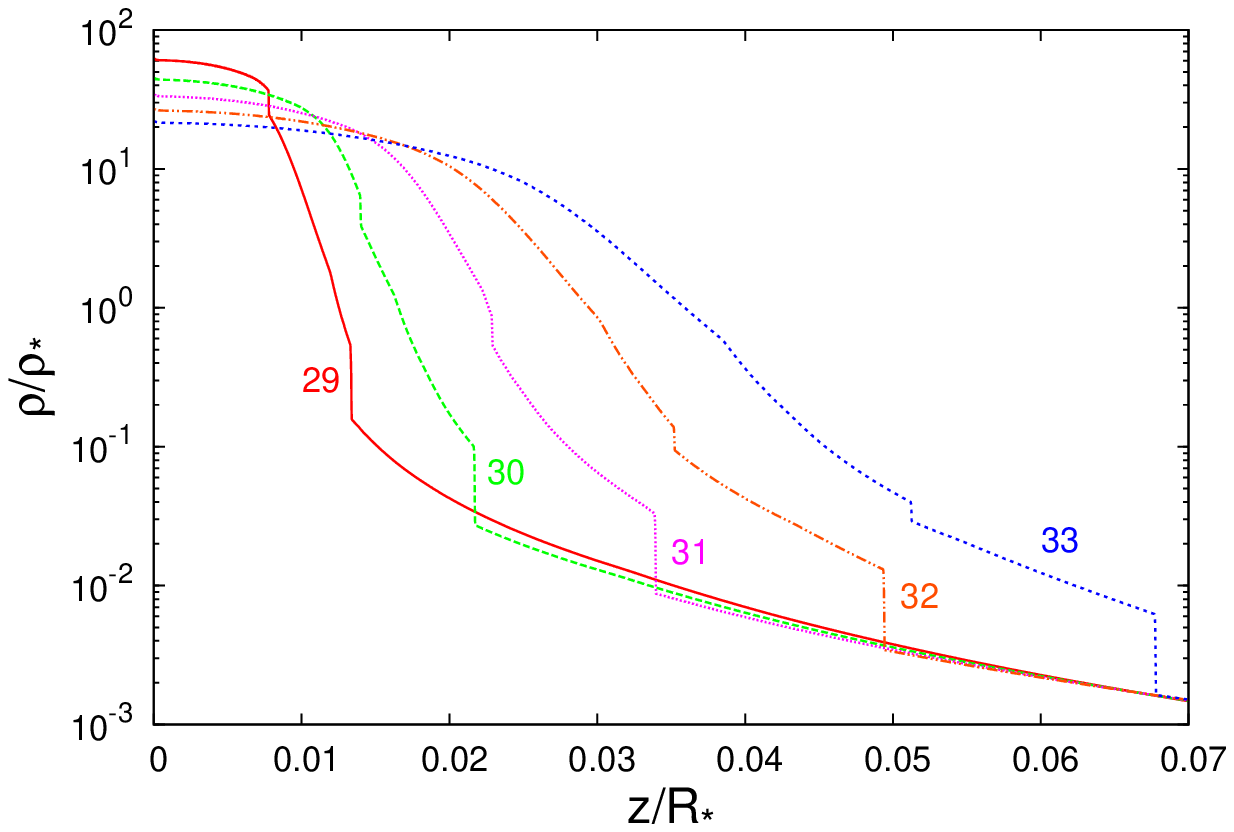} &
\hspace{-0.5cm}
(b) &
\hspace{-0.8cm}
\includegraphics[width=4.5cm]{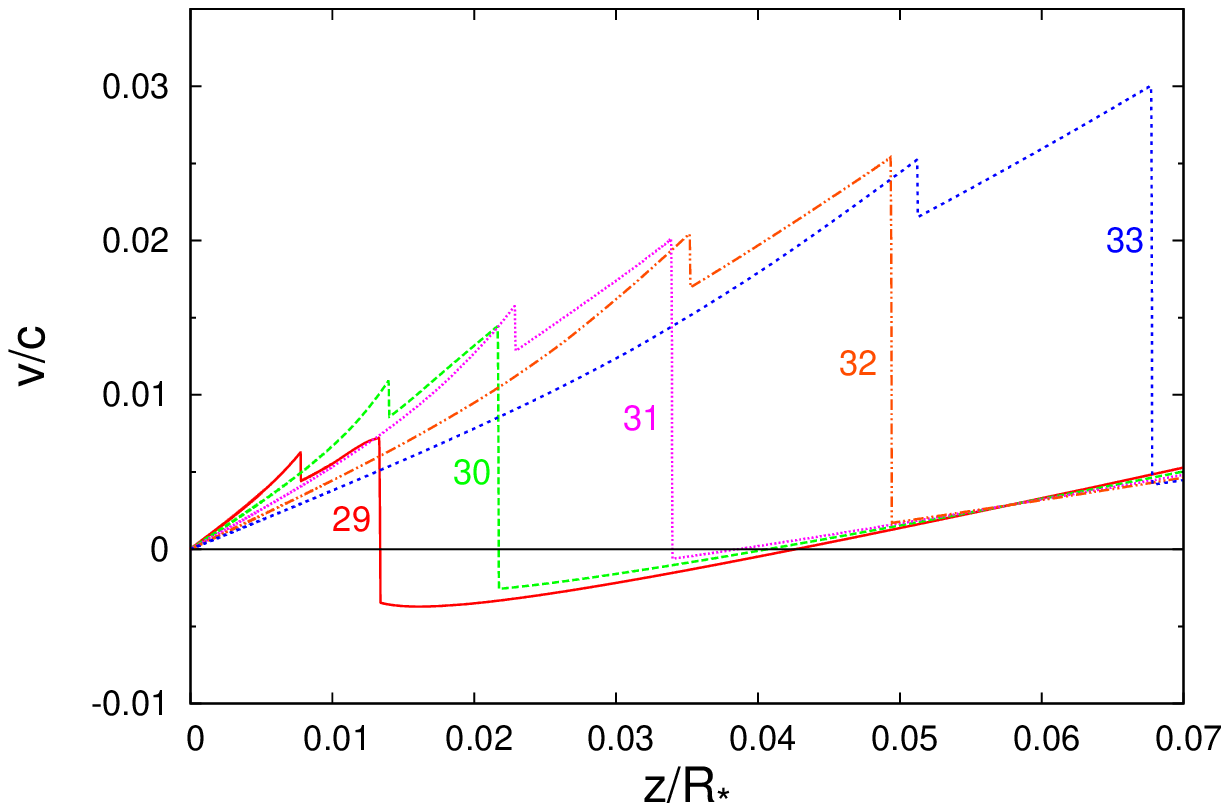}
\end{tabular}
\caption{
Density (left) and velocity (right) profiles in the positive vertical direction $z$ at different proper times $\tau$ during the second bounce-expansion phase for $\beta=9$.
The temporal evolution reads from top to bottom.
Labels stand for the following values of $\tau$ $[\mathrm{s}]$: (25)~89.51, (26)~89.98, (27)~90.45, (28)~90.94, (29)~91.38, (30)~92.54, (31)~93.72, (32)~94.89, (33)~96.05.
The star is at the periastron at $\tau=0$.
The point on the velocity profiles corresponds to the sonic point, the flow being subsonic (resp. supersonic) to the left (resp. right).
After its reflexion at the centre of the star at $\tau \approx 89.13$, the small-amplitude shock wave $S_{3}$ propagates outwards, in front of the stellar matter which begins to expand.
Pressure waves induced on the still collapsing stellar matter leads to the formation of another shock wave $S_{4}$ in front of the first one (a).
This shock wave produces a strong compression ratio $\approx 3.7$, and reaches the stellar surface at $\tau \approx 96.18$ (b).
It then occurs an over-all expansion.
}
\label{fig_g_08}
\end{figure}

The reflexion of both shock waves definitely stops the central compression, and the stellar matter begins again to expand from the centre (Fig.~\ref{fig_g_08}).
In front of the reflected shock wave, a last shock wave $S_{4}$ finally forms following the pressure waves which are generated by the opposite motions of expansion and collapse. 
This strong shock wave propagates up to the stellar surface and escapes followed by the previous one. 
The whole compressive motion stops to give way to a second phase of over-all expansion while the star moves towards the tidal radius.
\\
\\
\noindent
\textbf{Observational signature via $X$/$\gamma$ -ray bursts}

As already highlighted in \citetalias{bra2008}, the shock waves generated during the single Newtonian orthogonal compression could highly heat up the surface of the star in the hard $X$ or soft $\gamma$ -ray domains. 
Such a mechanism may provide a \textit{direct} observational signature of strongly disruptive encounters between stars and massive black holes in galactic nuclei.

\begin{figure}
\begin{center}
\begin{tabular}{cc}
\includegraphics[width=4.5cm]{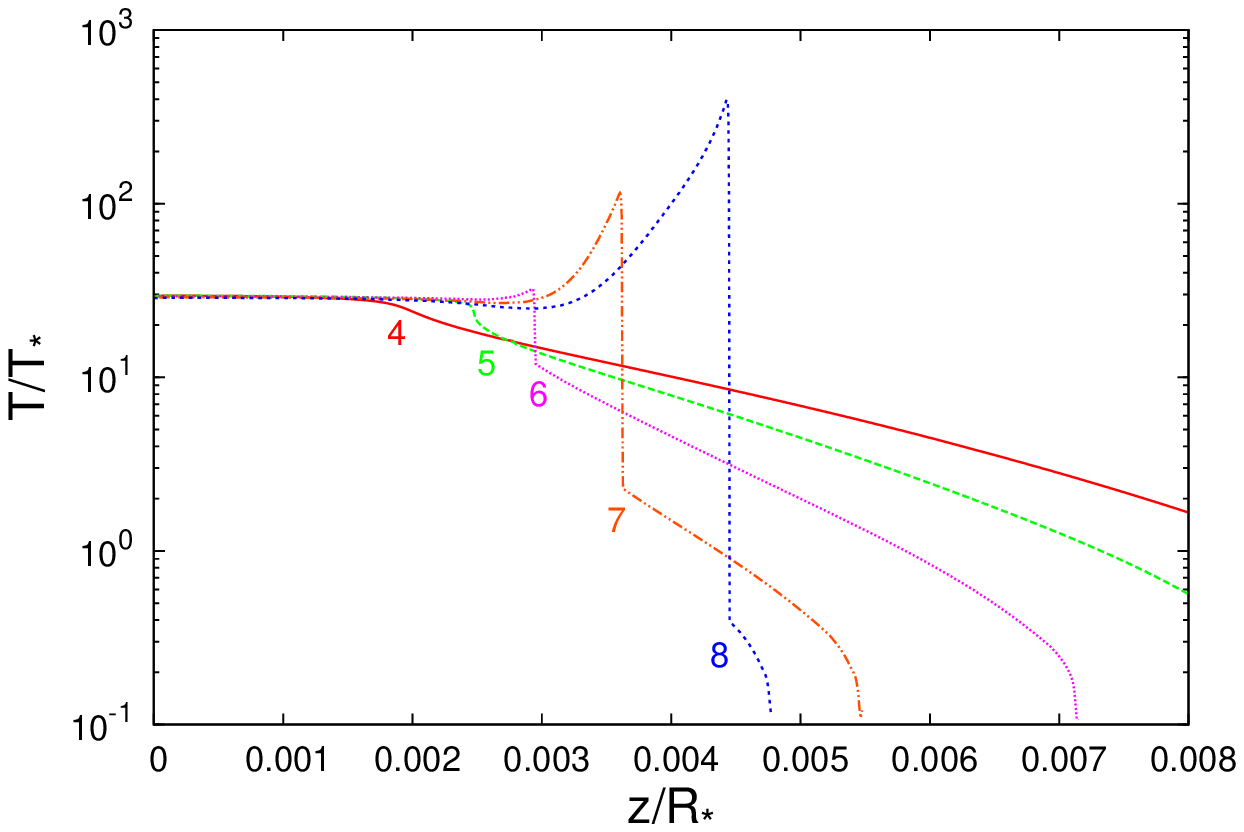} &
(a) \\
\includegraphics[width=4.5cm]{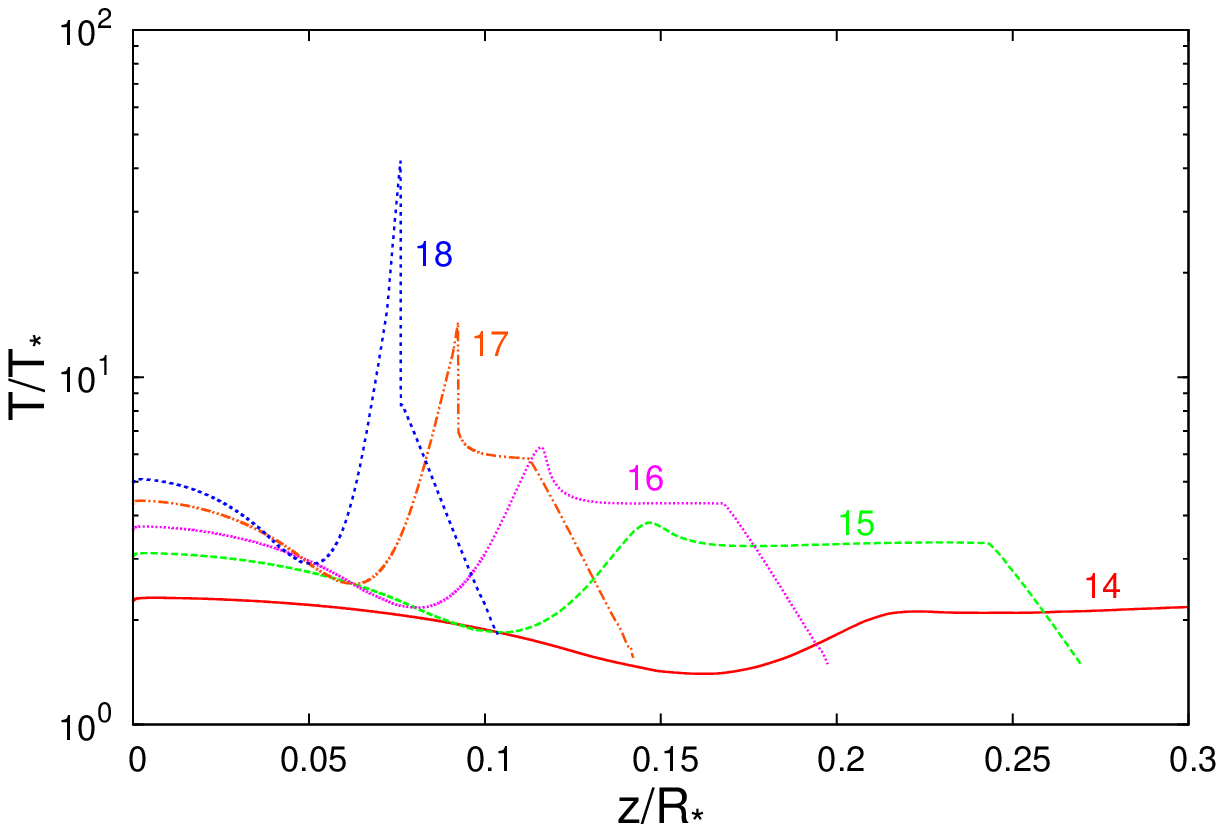} &
(b) \\ 
\includegraphics[width=4.5cm]{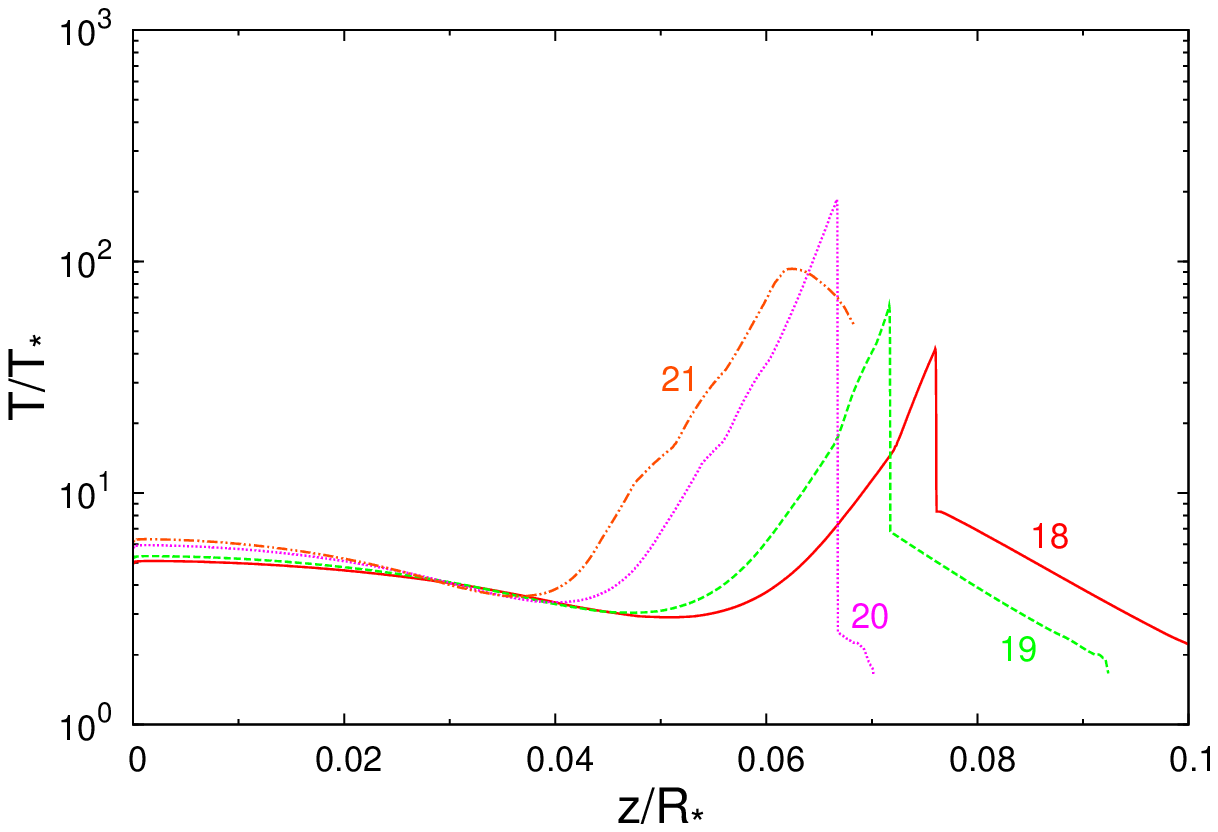} &
(c) \\
\includegraphics[width=4.5cm]{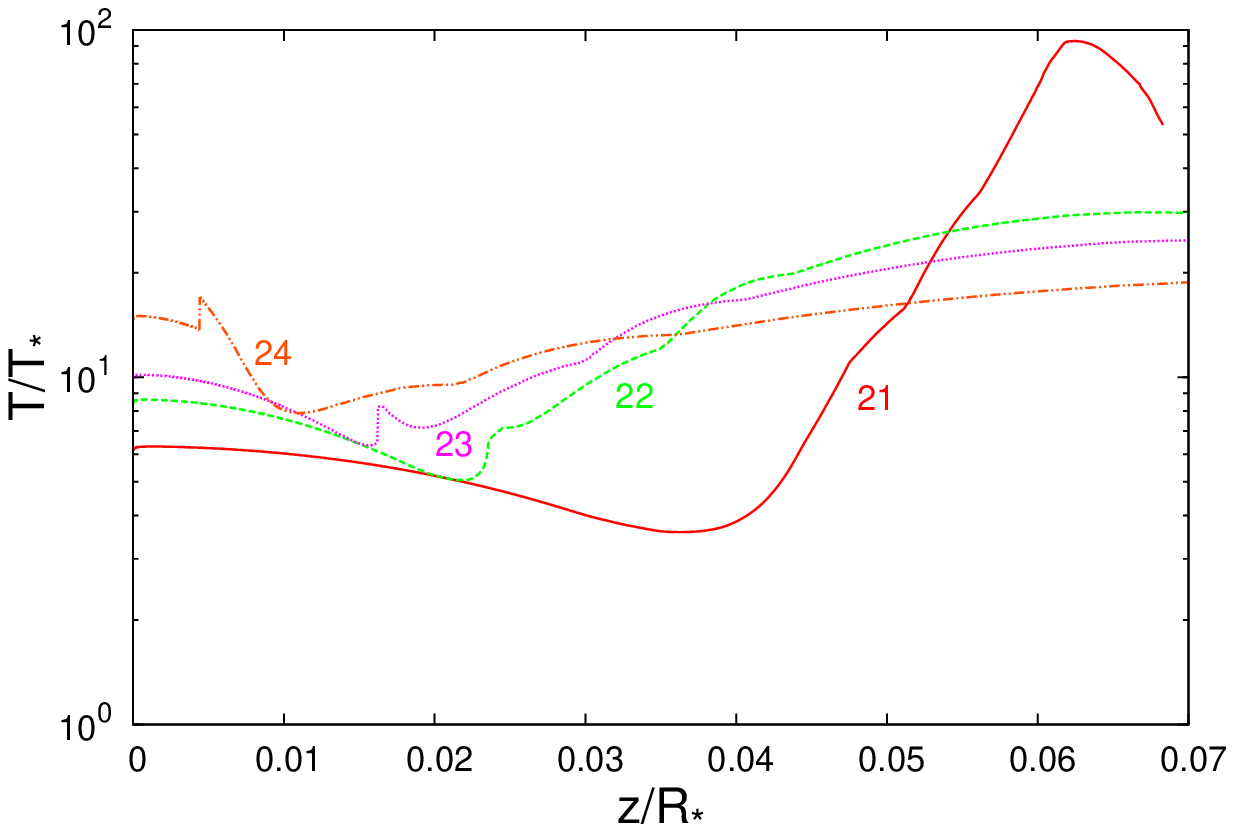} &
(d) \\
\includegraphics[width=4.5cm]{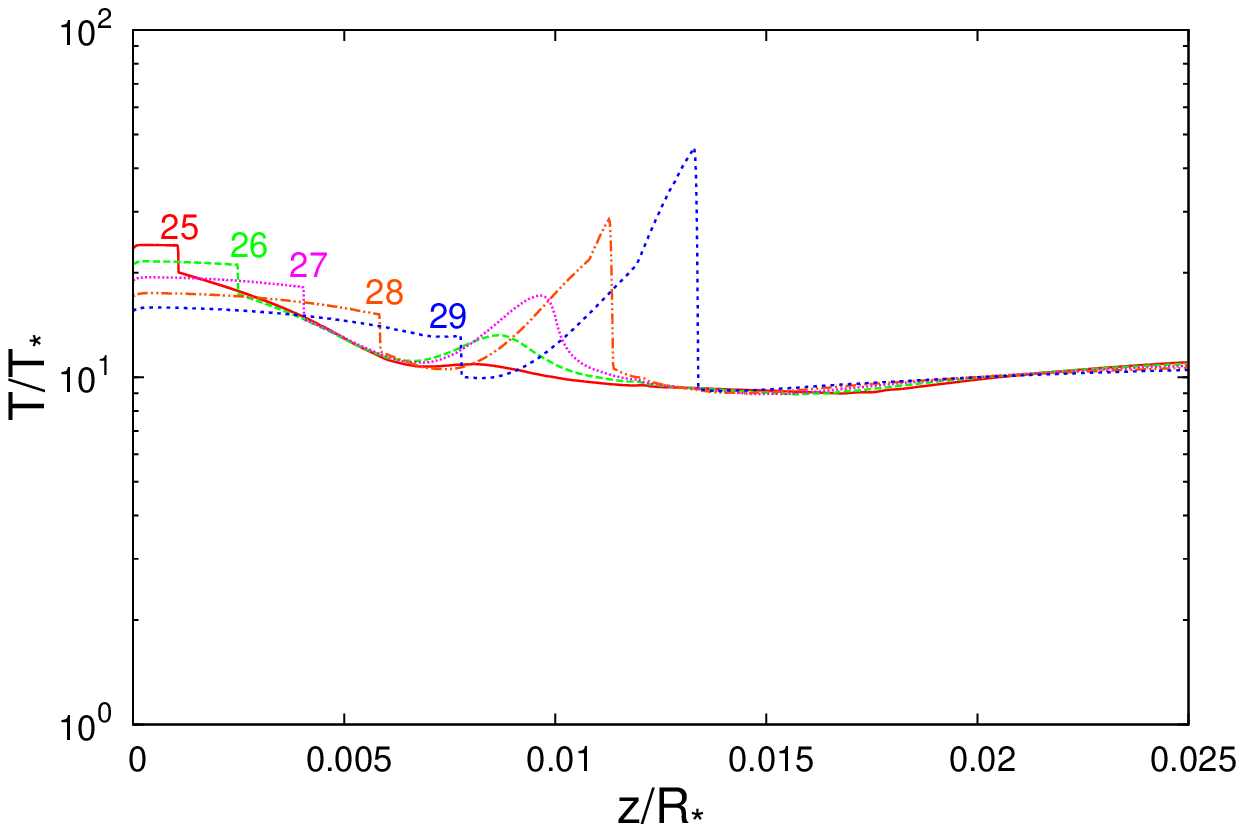} &
(e) \\
\includegraphics[width=4.5cm]{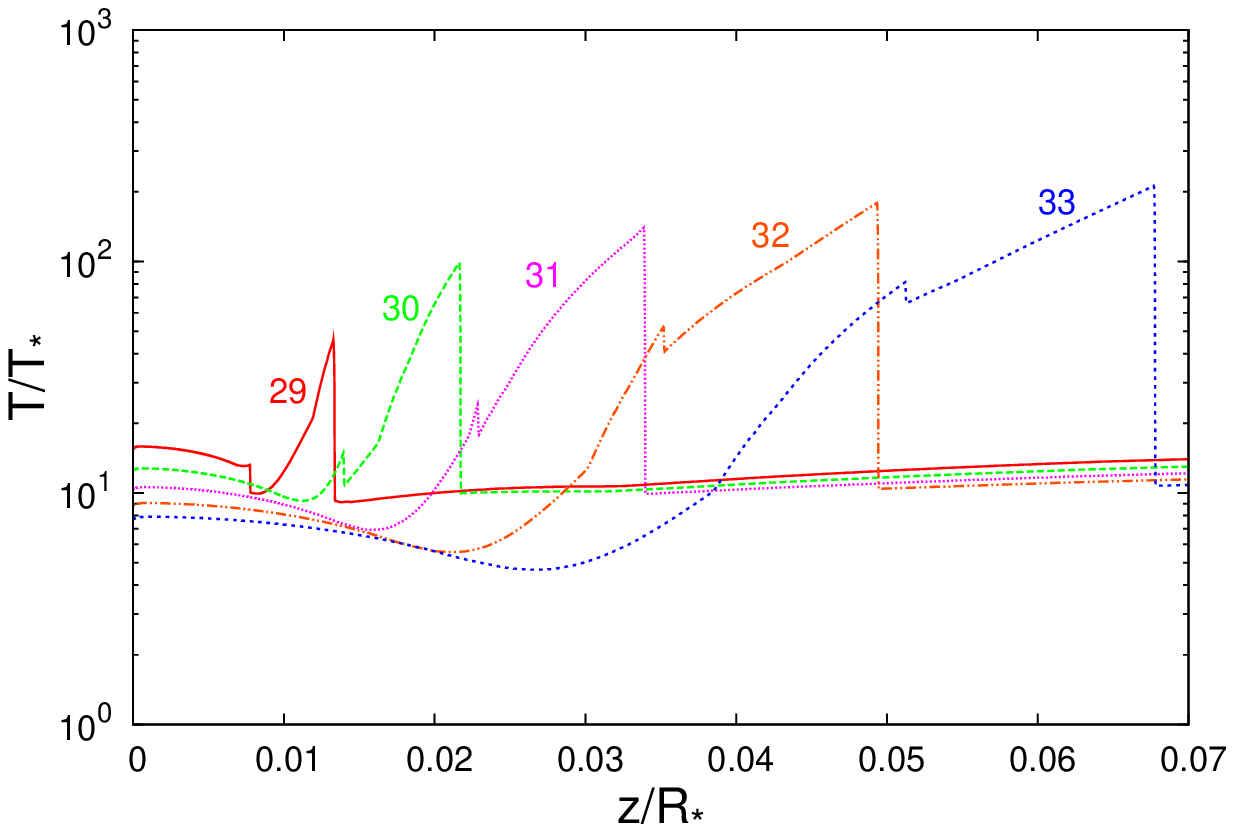} &
(f)
\end{tabular}
\caption{
Temperature profiles in the positive vertical direction $z$ at different proper times $\tau$ for $\beta=9$.
The temporal evolution reads from top to bottom.
Labels stand for the following values of $\tau$ $[\mathrm{s}]$: (4)~-37.68, (5)~-37.56, (6)~-37.33, (7)~-37.09, (8)~-37.00, (14)~56.48, (15)~68.13, (16)~72.80, (17)~76.29, (18)~78.61, (19)~79.31, (20)~80.69, (21)~81.40, (22)~84.31, (23)~85.49, (24)~87.86, (25)~89.51, (26)~89.98, (27)~90.45, (28)~90.94, (29)~91.38, (30)~92.54, (31)~93.72, (32)~94.89, (33)~96.05.
The star is at the periastron at $\tau=0$.
(a): first bounce-expansion phase (see Fig.~\ref{fig_g_04} (b)). 
The shock wave $S_{1}$ carries out a temperature jump from the centre to the surface of the star and escapes from the medium.
(b), (c), (d): second free-fall phase (see Fig.~\ref{fig_g_06}).
The shock wave $S_{2}$ carries out a temperature jump from the surface to the centre of the star, but is caught up by the collapsing surface and thus heats it up before escaping.
The shock wave $S_{3}$ then carries out a small temperature jump to the centre of the star.
(e), (f): second bounce-expansion phase (see Fig.~\ref{fig_g_08}).
The shock wave $S_{4}$ carries out a temperature jump from the centre to the surface of the star and escapes from the medium.
}
\label{fig_g_09}
\end{center}
\end{figure}

\begin{figure}[!h]
\resizebox{\hsize}{!}{\includegraphics{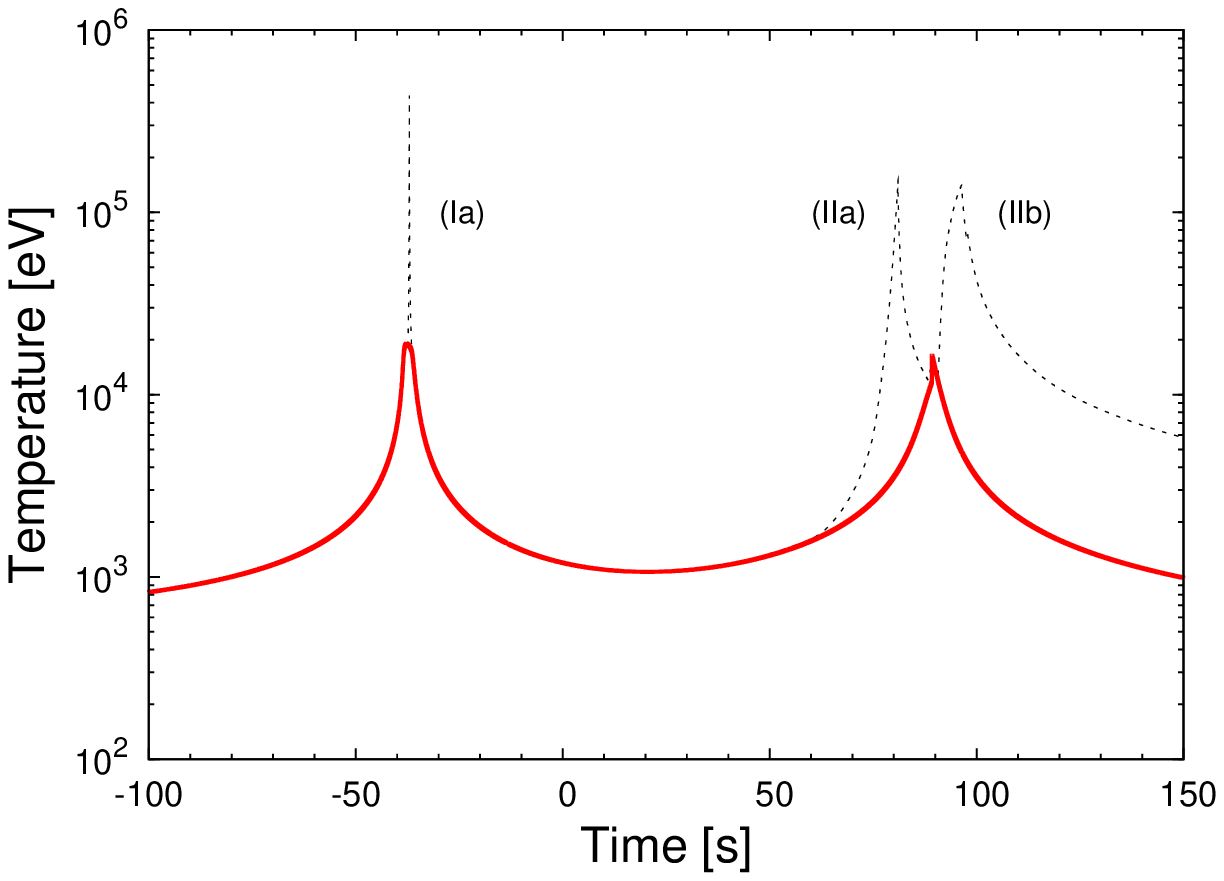}}
\caption{
Evolution of the temperature as a function of proper time $\tau$ for $\beta=9$.
The star is at the periastron at $\tau=0$.
The solid line corresponds to the temperature at the centre of the star.
The compression induces a first maximum $\approx 2 \times 10^{8} \, \mathrm{K}$ during $\approx 4 \, \mathrm{s}$, then a second one with the same amplitude during $\approx 6 \, \mathrm{s}$.
The dashed line corresponds to the temperature carried out along the vertical direction by the shock waves.
The three maxima are reached at the stellar surface.
(Ia): temperature increase due to the propagation from the centre to the surface of the shock wave $S_{1}$ resulting from the first bounce-expansion phase.
The surface reaches a temperature $\approx 4 \times 10^{9} \, \mathrm{K}$ during $\approx 0.03 \, \mathrm{s}$. 
(IIa): temperature increase due to the propagation from the surface to the centre of the shock wave $S_{2}$ resulting from the second free-fall phase.
Since it finally meets the shock wave, the stellar surface is heated up $\approx 10^{9} \, \mathrm{K}$ during $\approx 1.5 \, \mathrm{s}$.
(IIb): temperature increase due to the propagation from the centre to the surface of the shock wave $S_{3}$ resulting from the second bounce-expansion phase.
The surface reaches a temperature $\approx 10^{9} \, \mathrm{K}$ during $\approx 3 \, \mathrm{s}$. 
It lasts $\approx 115 \, \mathrm{s}$ between the first and second peaks of temperature at the surface, but a quite shorter time $\approx 15 \, \mathrm{s}$ between the second and the third ones.
}
\label{fig_g_10}
\end{figure}

The temperature profiles associated to the double relativistic orthogonal compression of the star are reproduced in Fig.~\ref{fig_g_09}.
The different shock waves actually lead to heat up three times the stellar surface.
A first time by the shock wave $S_{1}$ formed during the first bounce-expansion phase.
A second time by the shock wave $S_{2}$ formed during the second free-fall phase, and finally caught up by the faster stellar surface.
A third time, quite near to the previous one, by the shock wave $S_{4}$ formed during the second bounce-expansion phase.
At each occasion, the shock waves drop off a temperature $\approx 10^{9} \, \mathrm{K}$ at the stellar surface (Fig.\ref{fig_g_10}).
Therefore, it appears that the prompt emission of hard $X$ or soft $\gamma$ -ray bursts which may be associated to the stellar pancake mechanism would be interestingly intensified by the relativistic effects of multiple compressions.
%
%
%
%
\section{Conclusion}
\begin{figure}
\begin{center}
\begin{tabular}{c}
\includegraphics[width=8.5cm]{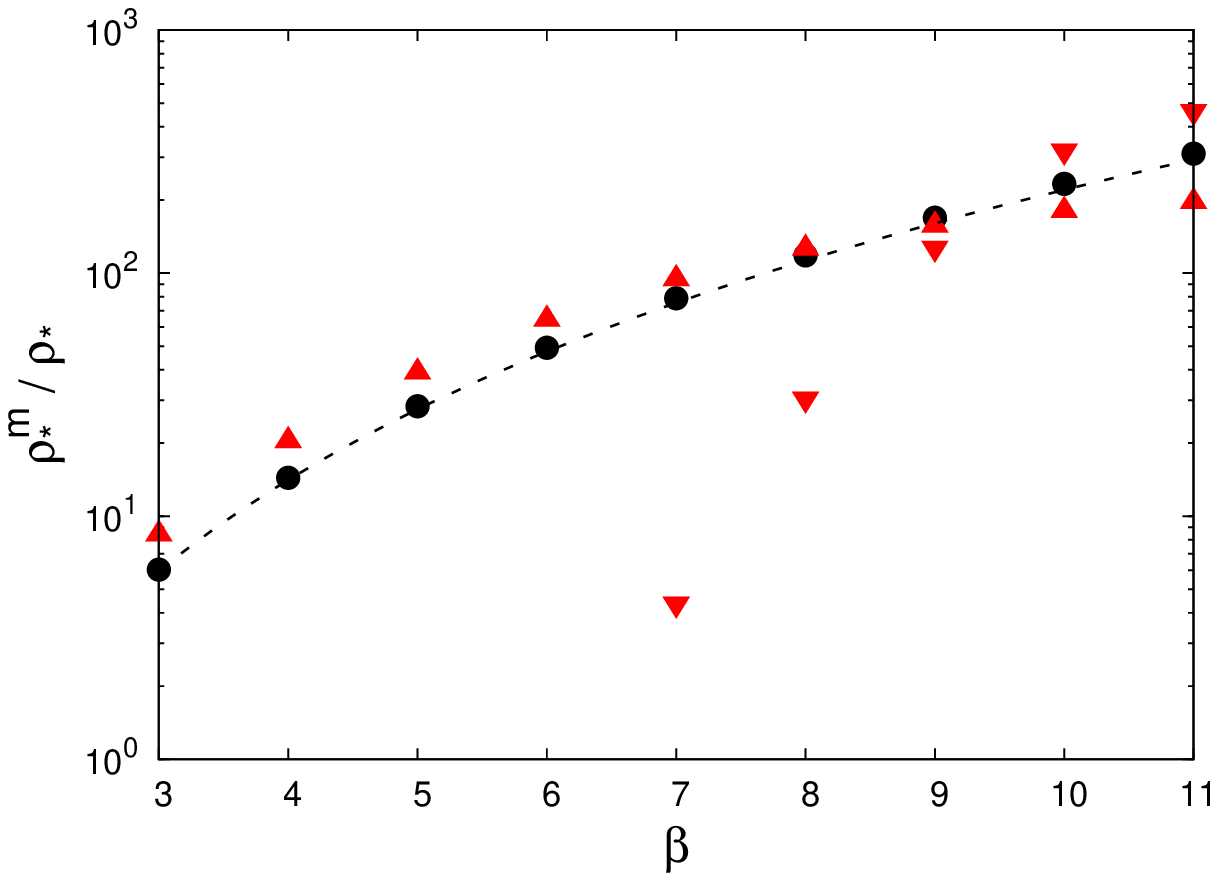} \\
\includegraphics[width=8.5cm]{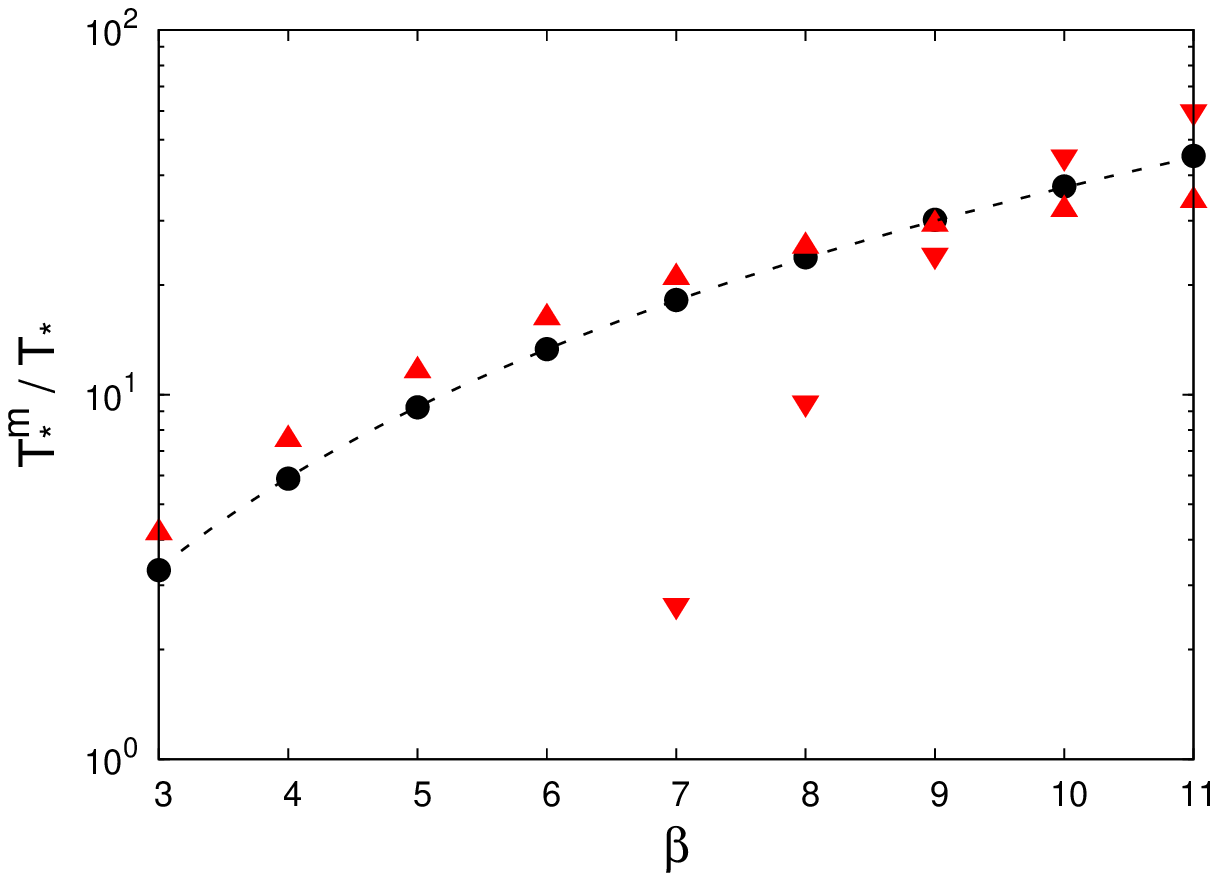} \\
\includegraphics[width=8.5cm]{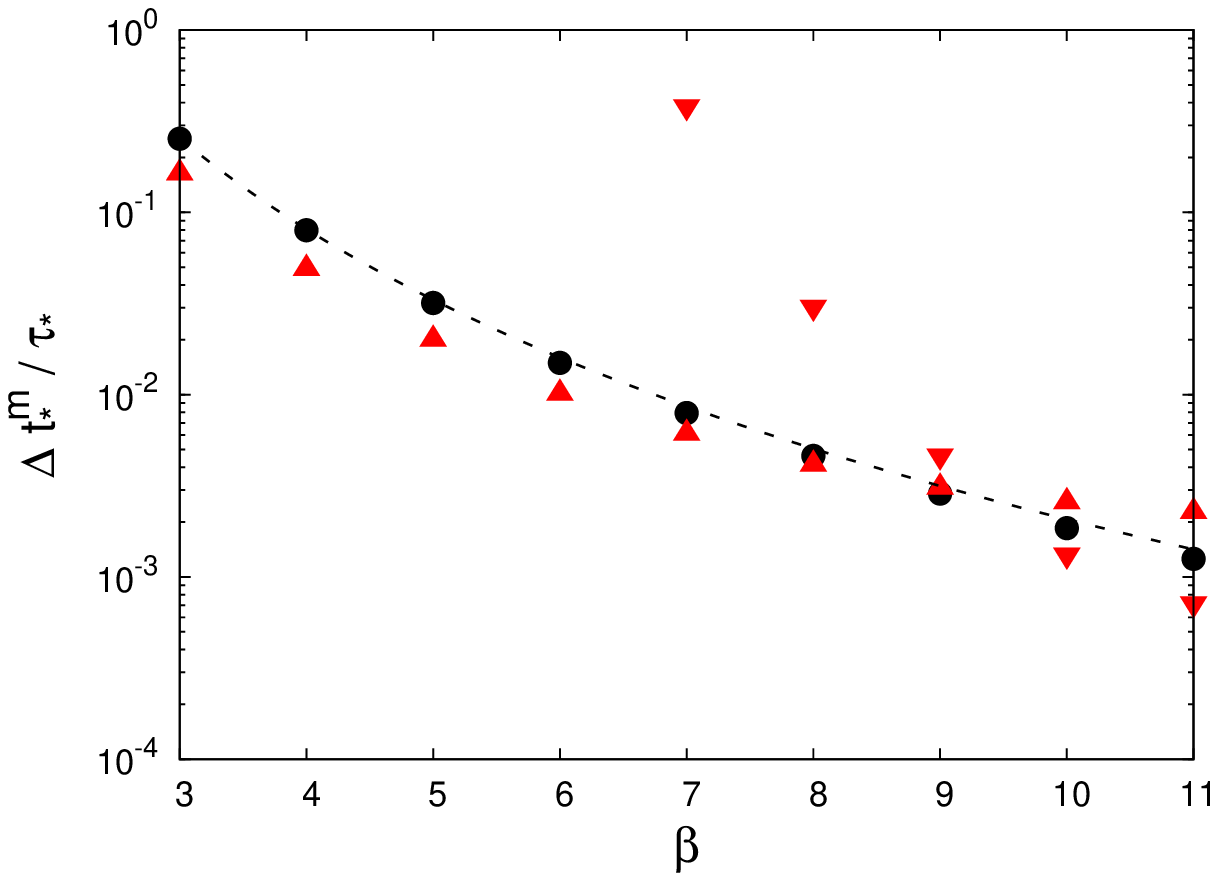}
\end{tabular}
\caption{
Evolutions of the maximum central density (top), maximum central temperature (middle), and duration during which the highest values are maintained (bottom) as functions of the penetration factor $\beta$ in the case $\gamma=5/3$.
The rightside up (resp. upside down) triangles indicate the first (resp. second) compression.
The points show for comparison the single compression calculated for a Newtonian gravitational field, and the dashed lines correspond to the power laws predicted by the affine model (\citealt{car1983}; see also \citetalias[][Eqs.~(39)-(41) and Fig. 16]{bra2008}).
In bottom figure, the characteristic internal timescale of the star $\tau_{\star}=(1/G\rho_{\star})^{1/2}$. 
}
\label{fig_i_01}
\end{center}
\end{figure}

We have presented first explorative results on the star's tidal compressions by a massive Schwarzschild black hole which take account of the intrinsic shock waves formation orthogonal to the orbital plane.
Whereas the stellar core would be heated up by the external compressive tidal field (Fig.~\ref{fig_i_01}), the hydrodynamical simulations also suggest that the stellar surface would be heated up during the shock waves propagation.

Besides the more restrictive scenario of the core thermonuclear explosion, it is thus believed that the close passage of a star to a massive black hole may be directly revealed as highly energetic short-timescale bursts originating from the shocked stellar surface. 
It is well known that some observed gamma-ray light curves present two or more peaks with complex structures \citep[e.g.][]{fis1995}.
Could some of them be interpreted as tidally-induced pancake stars? 
Typical timescales shown in Fig.~\ref{fig_g_10} for a star - black hole encounter along a relativistic self-crossing orbit are quite consistent with those of a short duration gamma-ray burst with two  peaks of energy separated by a hundred of seconds, comparable to what has been observed e.g. in GRB 970815 \citep{smi2002}.
The long duration GRB 060614 has also been tentatively interpreted by other authors as a strongly disruptive star - black hole encounter \citep{lu2008}.

Tidally-induced gamma-ray bursts are estimated to occur once every $10^{3}$--$10^{5}$ years per galaxy, depending on the nuclear stellar density profile and the mass of the central black hole \citep{wan2004}.
Such a rate can even be increased by a factor $10^{4}$ in the case of a massive binary black hole \citep{che2009}.
Since most of the galaxies --~including our own Milky Way~-- harbour a central massive black hole, and since the full observable universe is transparent in $X$ or $\gamma$ -ray domains, several events of this kind would be then detectable each year.

To well characterize this signature and make valuable comparisons with observational data of high-energy bursts, simulations have to consider the full stellar pancake's structure and the radiative transfer of energy through the layers of the star.
From this perspective, three-dimensional high-resolution shock-capturing simulations have been recently performed in the context of a Newtonian gravitational field, and have shown that shock waves actually form in several other directions within the star with the consequence of isotropizing the surface temperature \citep{gui2009}.
%
%
%
%
\begin{acknowledgements}
Matthieu Brassart warmly thanks the Fondation des Treilles (France) for the financial support of the present study.
\end{acknowledgements}
%
%
%
%

%
%
\end{document}